\documentclass[letterpaper,aps,prb,showpacs,intlimits,amsmath,amssymb,twocolumn,superscriptaddress]{revtex4}

\usepackage{graphicx}
\usepackage{bm}

\DeclareMathOperator{\Img}{Im}
\DeclareMathOperator{\Real}{Re}
\DeclareMathOperator{\Tr}{Tr}

\begin{document}

\title{Nonresonant Raman and inelastic X-ray scattering in the
charge-density-wave phase of the spinless Falicov-Kimball model}

\author{O.~P.~Matveev}
\affiliation{Institute for Condensed Matter Physics of
the National Academy of Sciences of Ukraine, Lviv, 79011 Ukraine}

\author{A.~M.~Shvaika}
\affiliation{Institute for Condensed Matter Physics of
the National Academy of Sciences of Ukraine, Lviv, 79011 Ukraine}

\author{J.~K.~Freericks}
\affiliation{Department of Physics, Georgetown University,
Washington, DC 20057, U.S.A.}

\begin{abstract}
The dynamical mean-field theory formalism to describe 
nonresonant inelastic light and X-ray scattering in a charge-density-wave phase is developed and applied to the spinless Falicov-Kimball model on an infinite-dimensional hypercubic lattice at half filling.  At zero temperature, the charge gap in the density of states is exactly equal to $U$; increasing the temperature rapidly fills the gap with subgap states. The nonresonant response function for Raman and inelastic X-ray scattering shows peaks connected with transitions over the gap and transitions that involve subgap states; in addition, the spectra have significant changes in shape as the temperature is raised from zero to $T_c$. In the case of X-ray scattering (when both energy and momentum are transferred), the response function illustrates features of dynamical screening (vertex corrections) in the different (nonresonant) symmetry channels ($A_{\rm 1g}$ and $B_{\rm 1g}$); dynamical screening is also present in the $A_{\rm 1g}$ Raman signal. Finally, we derive and verify the first moment sum rules for the (nonresonant) Raman and inelastic X-ray response functions in the charge-density-wave phase and we discuss experimental implications for how the sum rules might be employed in data analysis.

\end{abstract}


\maketitle

\section{Introduction}
Charge-density-wave (CDW) systems possess a static rearrangement of the charge that is modulated by their
ordering vector.  Since the underlying ionic cores are charged, they will respond to this charge modulation
from the electrons, and often create a distorted lattice structure that follows the modulated charge order
of the electrons.  This is often one of the easiest to measure signals of CDW order, namely the distortion of
the unit cell due to the ionic displacement that goes hand-in-hand with the electronic charge modulation; it is more difficult to directly measure the electronic charge modulation in the material.

In this work, we focus on signatures of the CDW order that are present in inelastic light scattering experiments
on CDW systems.  Since inelastic Raman scattering is sensitive to different symmetry charge modulations
(when polarizers are used on the incident and scattered light), it can provide information about the symmetry
of the CDW state which is complementary to the results that would come from an elastic light scattering measurement such as optical reflectivity (which can measure only one symmetry).  Similarly, because inelastic X-ray scattering also allows for an exchange of momentum by the scattered
photon, we might anticipate interesting behavior to occur when the ordering wave vector and the transferred momentum are
the same.

We develop all of the formalism to generalize the dynamical mean-field theory (DMFT) approach to inelastic Raman and
X-ray scattering in the situation when there is a CDW phase on a bipartite lattice with an ordering wave vector
equal to $(\pi,\pi,\ldots,\pi)$; our formulas include all effects of vertex corrections including dynamical screening.  While the formal development, in terms of the Green's functions, self-energies, and
irreducible vertex functions, is completely general, and can be applied to any many-body model that has CDW order, such as the attractive Hubbard model or the Holstein model, we analyze the formalism for the specific case of the Falicov-Kimball model because the irreducible charge vertex is known exactly, and so we can provide an exact solution to the light scattering problem.  In addition to deriving formulas for the light scattering spectra, we also examine the first moment sum rules for these spectra, which are equal to expectation values related to the kinetic and potential energies of the material. These sum rules can be employed to aid in the data analysis of experiments, when higher-energy bands are well separated from the low energy band that undergoes the CDW order, as already observed in systems that do not have CDW order, like in the normal state of SmB$_6$ at low temperature. They also provide an alternative way to directly measure the electronic order parameter of the CDW.

We anticipate our results should be relevant to different
experimental systems that display charge-density-wave order via nesting on a bipartite lattice at half filling,
especially in compounds which are three-dimensional
such as\cite{cdw_exp} BaBiO$_3$ and Ba$_{1-x}$K$_{x}$BiO$_3$, because DMFT is most accurate in higher dimensional systems; it may also be relevant to some layered two-dimensional systems, at least in a semi-quantitative fashion.
Our work also extends recent results on transport and optical conductivity in CDW systems\cite{krishnamurthy,matveev} to the realm of inelastic light scattering. Since inelastic light scattering experimental work on CDW systems has focused on Raman scattering of the soft phonon modes, the next step experimentally will likely be to examine the electronic scattering directly (either with Raman or with X-rays). Hence this work has the potential to be directly relevant to the next generation of experiments in this area.

The paper is organized as follows:  in Sec.~II, we derive the formalism for inelastic light scattering
in a symmetry broken phase including explicit expressions for Raman scattering, inelastic X-ray scattering, and their first moment sum rules; this formal development is appropriate for any many-body model of light scattering with local interactions.  In Sec.~III, we present our numerical results for the example case of the spinless Falicov-Kimball model and discuss what signatures are likely to be seen in experiment.  Our conclusions are presented in Sec.~IV.

\section{Formalism}

Since CDW ordering is a static order, it is often well described by static models such as the
Falicov-Kimball model.\cite{falicov_kimball}  This model was
introduced in 1969 to describe metal-insulator transitions in rare-earth compounds and transition-metal oxides.
Since then, it has been studied widely within the DMFT community, primarily because it is one of the simplest many-body problems that admits an exact solution\cite{brandt_mielsch1} (for a review see Ref.~\onlinecite{freericks_review}).
The Falicov-Kimball model has two kinds of particles: mobile electrons and localized electrons. Mobile electrons hop from site to site with  a hopping integral between nearest neighbors and they interact with the localized electrons when both sit on the same site (the interaction energy is $U$); we denote the mobile electron creation (annihilation) operator at site $i$ by $\hat d_i^\dagger$ ($\hat d_i^{}$) and the local electron creation (annihilation) operator at site $i$ by $\hat f_i^\dagger$ ($\hat f_i^{}$). The model has commensurate CDW order at half filling and this is the main property we exploit here. Brandt and Mielsch worked out the formalism for calculating the ordered-phase Green's functions\cite{brandt_mielsch2} shortly after Metzner and Vollhardt introduced the idea of the many-body problem simplifying in large dimensions.\cite{metzner_vollhardt} The CDW order parameter was shown to
display anomalous behavior at weak coupling,\cite{vandongen,chen_freericks} and higher-period ordered
phases have been examined on the Bethe lattice.\cite{freericks_swiss} Transport calculations in the commensurate CDW phase have also appeared recently.\cite{krishnamurthy,matveev} For concreteness, we will focus our attention in the formalism section on the Falicov-Kimball model, but the light scattering formulas have a wider range of applicability.

\subsection{DMFT for the CDW ordered phase}

The hypercubic lattice is a bipartite lattice, implying that it separates into two sublattices (called $A$ and $B$) with the hopping being nonzero only between the two sublattices.  In this case, the model will display commensurate (chessboard) CDW order when both the light and heavy particles are half-filled.  This CDW order corresponds to the situation where the average filling of the electrons remains uniform on each sublattice, but changes from one sublattice to another (it is commensurate because the lattice is bipartite here). We begin by writing the Falicov-Kimball model Hamiltonian as the sum of its local and nonlocal parts
\begin{equation}\label{eq: ham_def2}
  \mathcal{\hat{H}}=\sum_{ia}\mathcal{\hat{H}}_{i}^{a}-
  \sum_{ijab}t_{ij}^{ab}\hat{d}_{ia}^{\dag}\hat{d}_{jb}^{},
\end{equation}
where $i$ and $a=A$ or $B$ are the site and sublattice indexes, respectively, and $t_{ij}^{ab}$ is the hopping matrix, which is nonzero only between different sublattices ($t_{ij}^{AA}=t_{ij}^{BB}=0$). The local Hamiltonian is equal to
\begin{equation}\label{eq: ham_loc2}
  \mathcal{\hat{H}}_{i}^{a}=U\hat{n}_{id}^{a}\hat{n}_{if}^{a}-
  \mu_{d}^{a}\hat{n}_{id}^{a}-\mu_{f}^{a}\hat{n}_{if}^{a};
\end{equation}
with the number operators of the mobile and localized electrons given by $\hat n_{id}=\hat d_i^\dagger \hat d_i^{}$ and $\hat n_{if}=\hat f_i^\dagger\hat f_i^{}$, respectively. Note that we have introduced different chemical potentials for different sublattices.   This is convenient for computations, because it allows us to work with a
fixed order parameter, rather than iterating the DMFT equations to determine the order parameter (which is subject to critical slowing down near $T_c$).  Of course, the equilibrium solution occurs when the chemical potential is uniform throughout the system ($\mu^A_d=\mu^B_d$ and $\mu^A_f=\mu^B_f$).

We start with the definition of the lattice Green's function
\begin{align}\label{eq: green_def}
  G_{ij}^{ab}(\tau)&=-\Tr\left [
  \mathcal{T}_\tau e^{-\beta\mathcal{\hat H}} \hat d_{ia}^{}(\tau)
  \hat d^\dagger_{jb}(0)\right ] / \mathcal{Z},
  \\
  \mathcal{Z}&=\Tr\exp[-\beta\mathcal{\hat H}].
  \nonumber
\end{align}
Within a Feynman-diagram formalism, the Green's function satisfies Dyson's equation (which in fact is a compact form of the diagrammatic series)
\begin{equation}\label{eq: dyson}
  \sum_{lc}[(\omega+\mu^{a}_{d})\delta_{ac}\delta_{il}-\Sigma_{il}^{ac}(\omega)
  +t^{ac}_{il}] G_{lj}^{cb}(\omega)
  =\delta_{ij}\delta_{ab},
\end{equation}
where $\omega$ is a real frequency. The unperturbed band structure for the hypercubic lattice with nearest neighbor (NN) hopping satisfies
\begin{equation}
  \epsilon_{\bm k}=-\sum_{i-j}t_{ij}^{AB}\exp[i{\bf k} \cdot({\bf R}_{iA}-{\bf R}_{jB})]
  =-2t\sum\limits_{\alpha=1}^D\cos{k_{\alpha}a},
\end{equation}
where ${\bf R}_{iA}$ is a lattice vector for site $i$ on sublattice $A$ and $a$ is the lattice spacing (we set $a=1$).

The first step of DMFT is to scale\cite{metzner_vollhardt} the hopping matrix element as $t=t^*/2\sqrt{D}$ (we use $t^*=1$ as the unit of energy) and then take the limit of the infinite dimensions $D\to\infty$. The self-energy is then local:
\begin{equation}\label{eq: sigma_local}
  \Sigma_{ij}^{ab}(\omega)=\Sigma_{i}^{a}(\omega)\delta_{ij}\delta_{ab},
\end{equation}
and in the case of two sublattices has two values $\Sigma^{A}(\omega)$ and $\Sigma^{B}(\omega)$. Now, we can write the solution of the Dyson equation (in a momentum representation) in a matrix form
\begin{equation}\label{eq: matrix_dyson}
  {G}_{\bm k}(\omega)=\left[{z}(\omega)-{t}_{\bm k}\right]^{-1},
\end{equation}
where $z(\omega)$ and the hopping term $t_{\bm k}$ are represented by the following $2\times2$ matrices:
\begin{align}\label{eq: twobytwo}
  {z}(\omega)&=\left ( \begin{array}{cccc}
  \omega+\mu^{A}_{d}-\Sigma^{A}(\omega) & 0  \\
  0 & \omega+\mu^{B}_{d}-\Sigma^{B}(\omega) \\
  \end{array}\right ),
  \\
  {t}_{\bm k}&=\left (\begin{array}{cccc}
  0 & \epsilon_{\bm k}  \\
  \epsilon_{\bm k} & 0 \\
  \end{array}\right ).
  \nonumber
\end{align}
After substituting Eq.~(\ref{eq: twobytwo}) into Eq.~(\ref{eq: matrix_dyson}), we obtain three equations for the different Green's function components
\begin{align}
  G_{\bm k}^{AA}(\omega)&=
  \dfrac{\omega+\mu^{B}_{d}-\Sigma^{B}(\omega)}{\bar{Z}^{2}(\omega)-\epsilon^2_{\bm k}},
  \label{eq: gaa}\\
  G_{\bm k}^{BB}(\omega)&=
  \dfrac{\omega+\mu^{A}_{d}-\Sigma^{A}(\omega)}{\bar{Z}^{2}(\omega)-\epsilon^2_{\bm k}},
  \label{eq: gbb}\\
  G_{\bm k}^{AB}(\omega)&=G_{k}^{BA}(\omega)=
  \dfrac{\epsilon_{\bm k}}{\bar{Z}^{2}(\omega)-\epsilon^2_{\bm k}}
  \label{eq: gab}
\end{align}
with $\bar Z(\omega)$ defined by
\begin{equation}\label{eq: barz_def}
  \bar{Z}(\omega)=\sqrt{[\omega+\mu^{A}_{d}-\Sigma^{A}(\omega)][\omega+\mu^{B}_{d}-\Sigma^{B}(\omega)]}.
\end{equation}
These expressions agree with those of Brandt and Mielsch\cite{brandt_mielsch2} even though our notation is somewhat different from theirs.

The local Green's functions are now found to be
\begin{equation}\label{eq: g_loc_sublattice}
  G^{aa}(\omega)=\dfrac{1}{N}\sum\limits_{\bm k}G^{aa}_{\bm
  k}(\omega)=\dfrac{\omega+\mu^{b}_{d}-\Sigma^{b}(\omega)}{\bar{Z}(\omega)}
  F_\infty[\bar{Z}(\omega)],
\end{equation}
where
\begin{equation}\label{eq: Hilb_trans}
  F_\infty[\bar{Z}(\omega)]=\int d\epsilon
  \rho(\epsilon)\frac{1}{\bar{Z}(\omega)-\epsilon}
\end{equation}
is the Hilbert transform of the noninteracting density of states (DOS), which satisfies $\rho(\epsilon)=\exp(-\epsilon^{2}/t^{*2})/t^*\sqrt{\pi}$ for the infinite-dimensional hypercubic lattice.

The second step of DMFT is to map the lattice Green's function onto a local problem by means of the dynamical mean field. Since there are two sublattices, a dynamical mean field $\lambda^a(\omega)$ is introduced on each of them. As a result, the local lattice Green's function on each sublattice becomes:
\begin{equation}\label{eq: lambda}
  G^{aa}(\omega)=\frac{1}{\omega+\mu^a_d-\Sigma^a(\omega)-\lambda^a(\omega)}.
\end{equation}
The third equation that closes the system of equations for $G^{aa}(\omega)$, $\Sigma^a(\omega)$ and $\lambda^a(\omega)$ is obtained from the condition that the local Green's function can be defined as the
Green's function of an impurity with the same dynamical mean field $\lambda^{a}(\omega)$. Such a problem can be exactly solved and the result is equal to
\begin{equation}\label{eq: impurity}
  G^{aa}(\omega)=\frac{1-n_f^a}{\omega+\mu^a_d-\lambda^a(\omega)}
  + \frac{n_f^a}{\omega+\mu^a_d-U-\lambda^a(\omega)}.
\end{equation}
This last equation must be modified if one solves a different many-body model such as the Hubbard or Holstein model, as one needs to solve the relevant impurity problem for the model being considered; the remainder of the algorithm is identical for other models.

These equations are self-consistently solved numerically.
The iterative DMFT algorithm to calculate the lattice Green's function is as follows: we analytically continue all of the above formulas to the Matsubara frequency axis, because calculations along this axis are much more stable and converge faster than those on the real axis.  Then, for a fixed value of the order parameter $\Delta n_{f}=n_f^A-n_f^B$, one chooses $n_f^A$ and $n_f^B$ in such a way that $n_f^A+n_f^B=2n_{f}$ ($n_{f}={1}/{2}$ for half filling). With those fixed quantities, we now propose a guess for the self-energy on each sublattice, and then compute the local Green's function from Eqs.~(\ref{eq: barz_def}) and (\ref{eq: g_loc_sublattice}). Then we extract the dynamical mean field on each sublattice from Eq.~(\ref{eq: lambda}), and find the local Green's function for the impurity from Eq.~(\ref{eq: impurity}). This value is substituted into Eq.~(\ref{eq: lambda}) to calculate the new self-energy. This procedure is repeated until the Green's function converges and we can calculate the filling of the conduction electrons; the chemical potential for the conduction electrons is adjusted so that the average conduction electron filling is equal to one half. In order to find the correct equilibrium order parameter $\Delta n_{f}$ at the given temperature, one next calculates the chemical potentials for the $f$-electrons on each sublattice via
\begin{eqnarray}\label{eq: chem_pot_imag}
  \mu^a_f&=&-\frac{U}{2}-T\ln \frac{1-n_f^a}{n_f^a}\\
&-&T\sum_n\ln \left [ 1-\frac{U}{i\omega_n+\mu^a_d-\lambda^a(i\omega_n)}\right ],\nonumber
\end{eqnarray}
where we introduce the fermionic Matsubara frequencies $i\omega_n=i\pi T(2n+1)$. If these two chemical potentials are not equal, then the order parameter chosen initially is incorrect, and one needs to repeat the iterative loop with a new $\Delta n_{f}$ to eventually satisfy the equilibrium condition where $\mu_f^A-\mu_f^B=0$; when this condition is satisfied, then $\Delta n_{f}$ is the order parameter at that temperature. This algorithm, where the order parameter is fixed and we check for equilibrium by examining the chemical potentials, and then update the fixed order parameter to achieve the equilibrium solution, does not suffer from critical slowing down, which does occur if we instead fix the chemical potentials and iterate the equations until they converge. Generically, the DMFT equations can be solved with an order of magnitude less computer time than if we use this alternative approach. Finally, we repeat this iterative solution on the real axis, with the chemical potentials and fillings fixed at their now known values, which also is much more efficient than trying to do the entire calculation on the real axis. For more complicated models, one most likely will need to fix the chemical potential and iterate the equations (which will be subject to critical slowing down near $T_c$), because one cannot solve the impurity problem with a fixed order parameter anymore.  This does not create any serious problems, it just requires more computer time.

In Ref.~\onlinecite{matveev}, we already analyzed the evolution of the DOS in the CDW-ordered phase. We reiterate the main points which will be needed here. At $T=0$, a real gap develops of magnitude $U$ with square root singularities at the band edges. As the temperature increases, the system develops substantial subgap DOS which are thermally activated within the ordered phase.  Plots of the DOS can be found in Ref.~\onlinecite{matveev}. Note that the singular behavior occurs for one of the ``inner'' band edges on each sublattice, and that the subgap states develop very rapidly as the temperature rises.

\subsection{Nonresonant inelastic scattering}

Now we develop the formalism for nonresonant light scattering in the CDW phase. We start from the standard formula for the inelastic light scattering cross section derived by Shastry and Shraiman\cite{shastry_shraiman}
\begin{align}\label{eq:_R}
  R(\bm q,\Omega)&=2\pi\sum\limits_{i,f}
  \dfrac{e^{-\beta\varepsilon_{i}}}{\mathcal{Z}}
  \delta(\varepsilon_{f}-\varepsilon_{i}-\Omega)
  \\
  & \times \left|g(\bm k_{i})g(\bm k_{f})e_{\alpha}^{i}e_{\beta}^{f}
  \left\langle f\left|\hat{M}^{\alpha\beta}(\bm q)\right|i\right\rangle \right|^{2}.
  \nonumber
\end{align}
It describes the scattering of band electrons by photons with $\Omega=\omega_{i}-\omega_{f}$ and $\bm q=\bm k_{i}-\bm k_{f}$ being the transferred energy and momentum, respectively, $\bm e^{i(f)}$ is the polarization of the initial (final) states of the photons and $\varepsilon_{i(f)}$ denotes the electronic eigenstates. The quantity $g(\bm q)=(hc^{2}/V\omega_{\bm q})^{1/2}$ is called the ``scattering strength'' with $\omega_{\bm q}=c|\bm q|$. The scattering operator $\hat{M}(\bm q)$ is constructed from both the number current operator and the stress tensor which are equal to
\begin{equation}
  j_{\alpha}(\bm q)=\sum_{ab\bm k}
  \dfrac{\partial t_{ab}(\bm k)}{\partial
  k_{\alpha}}\hat{d}_{a}^{\dag}(\bm k+\bm q/2)\hat{d}_{b}(\bm k-\bm q/2)
\end{equation}
and
\begin{equation}\label{eq:_gamma_op}
  \gamma_{\alpha,\beta}(\bm q)=\sum\limits_{ab\bm k}
  \dfrac{\partial^{2} t_{ab}(\bm k)}{\partial k_{\alpha}\partial
  k_{\beta}}\hat{d}_{a}^{\dag}(\bm k+\bm q/2)\hat{d}_{b}(\bm k-\bm q/2),
\end{equation}
respectively (in models with spin, and additional sum over the $z$-component of spin is required). Here $t_{ab}(\bm k)$ are the components of the $2\times 2$ hopping matrix in Eq.~(\ref{eq: twobytwo}). The interaction of an electronic system with a weak external transverse electromagnetic field $\mathbf{A}$ is described by the  Hamiltonian
\begin{align}
  H_{\textrm{int}}&=-\dfrac{e}{\hbar c}
  \sum_{\bm k}\bm j(\bm k)\cdot\bm A(-\bm k)
  \\
  &+\dfrac{e^{2}}{2\hbar^{2}c^{2}}
  \sum_{\bm k \bm k'}A_{\alpha}(-\bm k)\gamma_{\alpha,\beta}(\bm k+\bm k')A_{\beta}(-\bm k').
  \nonumber
\end{align}
The scattering operator $M$ is then constructed from these interaction terms; it has both nonresonant and resonant contributions
\begin{align}
  \left\langle f\left|\hat{M}^{\alpha\beta}(\bm q)\right|i\right\rangle
  &=\left\langle f\left|\gamma_{\alpha,\beta}(\bm q)\right|i\right\rangle
  \\
  &+\sum_{l}\Biggl(\dfrac{\left\langle f\left|j_{\beta}(\bm k_{f})
  \right|l\right\rangle  \left\langle l\left|j_{\alpha}(-\bm k_{i})
  \right|i\right\rangle }{\varepsilon_{l}-\varepsilon_{i}-\omega_{i}}
  \nonumber \\
  &+\dfrac{\left\langle f\left|j_{\alpha}(-\bm k_{i})
  \right|l\right\rangle  \left\langle l\left|j_{\beta}(\bm k_{f})
  \right|i\right\rangle }{\varepsilon_{l}-\varepsilon_{i}+\omega_{f}}
  \Biggr)
  \nonumber
\end{align}
with the sum $l$ over intermediate states, and after substituting into the cross section formula, one obtains three terms: a pure resonant term; a nonresonant term; and a mixed term (because it is constructed from the square of the scattering operator).

The nonresonant contribution is
\begin{align}\label{eq: nonres_cros}
  R_{N}(\bm q,\Omega)
  &=2\pi g^{2}(\bm k_{i})g^{2}(\bm k_{f})
  \\
  &\times\sum_{i,f}\dfrac{\exp(-\beta\varepsilon_{i})}{\mathcal{Z}}
  \tilde{\gamma}_{i,f}\tilde{\gamma}_{f,i}
  \delta(\varepsilon_{f}-\varepsilon_{i}-\Omega).
  \nonumber
\end{align}
The tilde denotes contractions with the polarization vectors:
\begin{equation}\label{eq: gamma_op_tilda}
  \tilde{\gamma}=\sum\limits_{\alpha\beta}e_{\alpha}^{i}\gamma_{\alpha,\beta}(\bm q)e_{\beta}^{f}
\end{equation}
with the notation $\mathcal{O}_{i,f}=\left\langle i\left|\mathcal{O}\right|f\right\rangle $ for the matrix elements of an operator $\mathcal{O}$. (Resonant and mixed diagrams will be examined elsewhere.)

The next step is to evaluate the summations in Eq.~(\ref{eq: nonres_cros}) via Green's function techniques. In general, such a procedure is nontrivial. But for the nonresonant contribution it is relatively straightforward\cite{freericks_deveraux,sh_v_f_d}. We start from the Matsubara function built on two time dependent stress-tensor operators
\begin{equation}
  \chi_{\tilde{\gamma},\tilde{\gamma}}(\tau- \tau^\prime)
  =\Tr\left [  \mathcal{T}_\tau e^{-\beta\mathcal{\hat H}}
  \tilde{\gamma}(\tau) \tilde{\gamma}(\tau^\prime)  \right ] / \mathcal{Z}.
\end{equation}
The imaginary time dependence of the stress-tensor operator is evolved (in the Heisenberg representation) with respect to the equilibrium Hamiltonian because this is a linear-response calculation. The symbol $\mathcal{T}_{\tau}$ is a time ordering operator. Further, we perform a Fourier transformation to the imaginary Matsubara frequencies. In thermal equilibrium, the two-particle correlation function depends only on the difference of the two time variables and our Matsubara frequency dependent function can be evaluated as
\begin{align}
  \chi_{\tilde{\gamma},\tilde{\gamma}}( i\nu)
  &=\sum_{i,f}\dfrac{\exp(-\beta\varepsilon_{i})}{\mathcal{Z}}
  \dfrac{\tilde{\gamma}_{i,f}\tilde{\gamma}_{f,i}}
  {\varepsilon_{f}-\varepsilon_{i}-i\nu}
  \\
  &\times\left[1-\exp(\beta(\varepsilon_{i}-\varepsilon_{f}))\right].
  \nonumber
\end{align}
Performing an analytic continuation to the real axis $i\nu\to\Omega\pm i0^{+}$ produces the known expression
\begin{eqnarray}
  R_{N}(\bm q,\Omega)=\frac{2\pi
  g^{2}(\bm k_{i})g^{2}(\bm k_{f})}{1-\exp(-\beta\Omega)}\chi_{N}(\bm q,\Omega),
\end{eqnarray}
where we introduced the nonresonant response function
\begin{equation}
  \chi_{N}(\bm q,\Omega)=\dfrac{1}{\pi}
  \Img\chi_{\tilde{\gamma},\tilde{\gamma}}(\Omega+i0^{+}).
\end{equation}

\begin{figure}[htb]
   \includegraphics[scale=0.4]{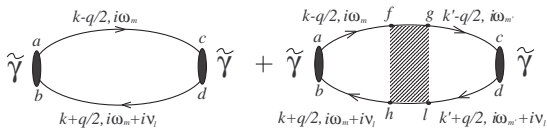}
   \caption{The Feynman diagrams for the generalized polarizations. Due to the properties of the irreducible
charge vertex of the Falicov-Kimball model, we will have $m=m^\prime$.}
   \label{fig: response}
\end{figure}

Now we have reduced the problem to that of finding the response function built on two stress-tensor
operators. Actually, such a function corresponds to a two-particle Green's function that will be shortly presented in Feynman diagrammatic notation. The Fourier transform of the two stress-tensor
correlation function can be represented as a sum over Matsubara frequencies of the ``generalized polarizations''
\begin{equation}
  \chi_{\tilde{\gamma},\tilde{\gamma}}(i\nu_l)=T\sum\limits_{m}\Pi_{m,m+l},
  \label{Pi}
\end{equation}
where we introduced the shorthand notation $\Pi_{m,m+l}=\Pi(i\omega_{m},i\omega_{m}+i\nu_{l})$ for the
dependence on the fermionic $i\omega_{m}=i\pi T(2m+1)$ and bosonic $i\nu_{l}=i2\pi Tl$ Matsubara frequencies. In the case of the CDW ordered phase, the Feynman diagrams for the ``generalized polarizations'' $\Pi_{m,m+l}$ are shown in Fig.~\ref{fig: response}, where we introduce additional sublattice indices $a$ to $l$.
Here, we used the fact that the total reducible charge vertex (shaded rectangle in Fig.~\ref{fig: response}) is a diagonal function of frequencies for the Falicov-Kimball model [see Eq.~(\ref{eq: Gamma}) below]; for other models, where the vertex is almost certainly no longer diagonal, the analysis is somewhat more complicated. Now one can perform an analytic continuation to the real axis and replace the sum over Matsubara frequencies by an integral over the real axis:
\begin{align}
  \chi_{\tilde{\gamma}\tilde{\gamma}}(i\nu_l)
  &=\dfrac{1}{2\pi i}\int_{-\infty}^{+\infty} d \omega f(\omega)
  \\
  &\times\Bigl{[}\Pi(\omega-i0^{+},
  \omega+i\nu_l)-\Pi(\omega+i0^{+},
  \omega+i\nu_l)
  \nonumber \\
  &+\Pi(\omega-i\nu_l,
  \omega-i0^{+})-\Pi(\omega+i\nu_l,\omega+i0^{+})\Bigr{]},
  \nonumber
\end{align}
where $f(\omega)=1\left/[\exp(\beta\omega)+1]\right.$ is the Fermi distribution function. Then the nonresonant response function is expressed directly in terms of the generalized polarizations
\begin{align}\label{eq: resp_nonres}
  \chi_{N}(q,\Omega)
  &=\dfrac{2}{(2\pi i)^{2}}\int\limits_{-\infty}^{+\infty}d\omega
  \left[f(\omega)-f(\omega+\Omega)\right]
  \\
  & \times
  \text{Re}\Bigl\{\Pi(\omega-i0^{+},\omega+\Omega+i0^{+})
  \nonumber \\
  &-\Pi(\omega-i0^{+},\omega+\Omega-i0^{+})\Bigr\}.
  \nonumber
\end{align}

The next step is to calculate these generalized polarizations. We consider both cases of inelastic light (Raman) and inelastic X-ray scattering. For Raman scattering, we can approximate $\bm q=0$ because the optical photon wavelength is so large, whereas for inelastic X-ray scattering, the transferred momentum is nonzero $\bm q\ne0$.

\subsection{Raman scattering: $\bm q=0$}

The non-resonant Raman response function presented in terms of the generalized polarizations in Eq.~(\ref{eq: resp_nonres}) is reduced to the calculation of the Feynman diagrams in Fig.~\ref{fig: response}. As a result, our aim is to calculate the sum of the products of the one-particle Green functions calculated in DMFT and the charge vertices. Here, the momentum ${\bm k}$ enters not only through the band energy term $\epsilon_{\bm k}$ [see Eqs.~(\ref{eq: gaa}-\ref{eq: gab})] but also through the stress-tensor factors, namely the derivatives $\partial^{2}\epsilon(\bm k)/\partial k_{\alpha}\partial k_{\beta}$. Furthermore, the stress-tensor operator is contracted with polarization vectors $\bm e^{i,f}$, see Eq.~(\ref{eq: gamma_op_tilda}), which vary for the different symmetries.

There are three symmetries often examined in experimental systems with cubic symmetry. The $A_{\rm 1g}$ symmetry has the full symmetry of the lattice and for the hypercubic lattice the incident and scattered light are both polarized along the same diagonal direction, so in large dimensions we take the initial and final polarizations to be
$\bm e^{i}=\bm e^{f}=(1,1,1,1,\ldots)$. The stress-tensor amplitude in the case of $A_{\rm 1g}$ symmetry (for NN hopping) is equal to minus the band energy
\begin{align}
  \bar{\gamma}_{A_{1g}}(\bm k)&=
  \sum_{\alpha\beta}e_{\alpha}^{i}
  e_{\beta}^{f}\dfrac{\partial^{2}\epsilon(\bm k)}{\partial k_{\alpha}\partial k_{\beta}}
=\dfrac{t^{*}}{\sqrt{D}}\sum_{\alpha=1}^D\cos{k_{\alpha}}=-\epsilon(\bm k)
\end{align}
The $B_{\rm 1g}$ symmetry is a $d$-wave-like symmetry that involves crossed polarizers along the diagonals. In this case, we take $\bm e^{i}=(1,1,1,1,\ldots)$ and $\bm e^{f}=(-1,1,-1,1,\ldots)$, so the stress-tensor amplitude is as follows
\begin{align}\label{eq: b1g_gamma}
  \bar{\gamma}_{B_{\rm 1g}}(\bm k)&=
  \sum_{\alpha\beta}e_{\alpha}^{i}
  e_{\beta}^{f}\dfrac{\partial^{2}\epsilon(\bm k)}{\partial
  k_{\alpha}\partial k_{\beta}}
  =-\dfrac{t^{*}}{\sqrt{D}}\sum_{\alpha=1}^{D}(-1)^{\alpha}\cos{k_{\alpha}}
\end{align}
Finally, the $B_{\rm 2g}$ symmetry is another $d$-wave-like symmetry rotated by 45 degrees; it requires the polarization vectors to satisfy $\bm e^{i}=(1,0,1,0,\ldots)$ and $\bm e^{f}=(0,1,0,1,\ldots)$, and for NN hopping there are no contributions to the nonresonant response in this channel.

We start with the analysis of the $B_{\rm 1g}$ symmetry, which is simplest case to examine. Here, the response function is determined only by the first term (bare loop) of the Feynman diagrams in Fig.~\ref{fig: response} and there are no  contributions from the second one\cite{khurana,freericks_deveraux} because the stress-tensor factor has momentum dependence that integrates to zero when multiplied by the local charge vertex and summed over all momentum.
\begin{figure}[htb]
  \includegraphics[scale=1]{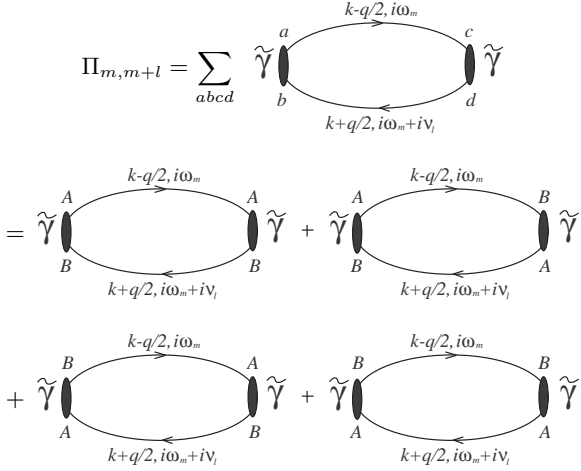}
  \caption{Individual terms for the bare polarization in the ordered
  phase. }
  \label{fig: polarization}
\end{figure}
The expanded form of the diagrams for the generalized polarization in the $B_{\rm 1g}$ channel for the CDW chessboard phase is presented in Fig.~\ref{fig: polarization} and is equal to
  \begin{align}\label{eq: sumP1}
  \Pi_{m,m+l}
  &=\frac{1}{N}\sum_{\bm k} \bar{\gamma}_{\bm k}^{2}
  \biggl(G_{\bm k-\tfrac{\bm q}2, m}^{AA}G_{\bm k+\tfrac{\bm q}2, m+l}^{BB}
  +G_{\bm k-\tfrac{\bm q}2, m}^{AB}G_{\bm k+\tfrac{\bm q}2, m+l}^{AB}
  \\
  &
   +G_{\bm k-\tfrac{\bm q}2, m}^{BA}G_{\bm k+\tfrac{\bm q}2, m+l}^{BA}
   +G_{\bm k-\tfrac{\bm q}2, m}^{BB}G_{\bm k+\tfrac{\bm q}2, m+l}^{AA}\biggr).
  \nonumber
  \end{align}
After substituting in the expressions for the Green's function in Eqs.~(\ref{eq: gaa}), (\ref{eq: gbb}), and (\ref{eq: gab}), and the expressions for the $\bar{\gamma}_{\bm k}$ amplitude from Eq.~(\ref{eq: b1g_gamma}), the individual contributions to $\Pi_{m,m+l}$ at $\bm q=0$ become
\begin{align}
  &\dfrac{1}{N}\sum\limits_{\bm k} \bar{\gamma}_{\bm k}^{2} G_{\bm k,m}^{AA}G_{\bm k,m+l}^{BB}
  =\dfrac{1}{2}(i\omega_m+\mu_{d}^{B}-\Sigma_{m}^B)
  \\
  &\times(i\omega_m+i\nu_l+\mu_{d}^{A}-\Sigma_{m+l}^A)
  \dfrac{\dfrac{F_\infty(\bar{Z}_{m+l})}{\bar{Z}_{m+l}}
  -\dfrac{F_\infty(\bar{Z}_m)}{\bar{Z}_m}}{\bar{Z}^{2}_m-\bar{Z}^{2}_{m+l}},
  \nonumber
\end{align}
\begin{align}
  &\dfrac{1}{N}\sum\limits_{\bm k} \bar{\gamma}_{\bm k}^{2} G_{\bm k,m}^{BB}G_{\bm k,m+l}^{AA}
  =\dfrac{1}{2}(i\omega_m+\mu_{d}^{A}-\Sigma_{m}^A)
  \\
  &\times(i\omega_m+i\nu_l+\mu_{d}^{B}-\Sigma_{m+l}^B)
  \dfrac{\dfrac{F_\infty(\bar{Z}_{m+l})}{\bar{Z}_{m+l}}
  -\dfrac{F_\infty(\bar{Z}_m)}{\bar{Z}_m}}{\bar{Z}^2_m-\bar{Z}^{2}_{m+l}},
  \nonumber
\end{align}
and
\begin{align}
  &\dfrac{1}{N}\sum\limits_{\bm k} \bar{\gamma}_{\bm k}^{2} G_{\bm k,m}^{AB}G_{\bm k,m+l}^{AB}
  =\dfrac{1}{N}\sum\limits_{\bm k} \bar{\gamma}_{\bm k}^{2} G_{\bm k,m}^{BA}G_{\bm k,m+l}^{BA}
  \\
  &=\dfrac{1}{2}\dfrac{\bar{Z}_{m+l}F_\infty(\bar{Z}_{m+l})-\bar{Z}_m F_\infty(\bar{Z}_m)}
  {\bar{Z}^{2}_m-\bar{Z}^{2}_{m+l}}.
  \nonumber
\end{align}
Hence, the total expression for the generalized polarization $\Pi_{m,m+l}$ is
\begin{align}\label{b1gP1}
  \Pi_{m,m+l}&=\dfrac{1}{2}\Biggl\{
  \frac{\dfrac{F_\infty(\bar{Z}_{m+l})}{\bar{Z}_{m+l}}
  -\dfrac{F_\infty(\bar{Z}_{m})}{\bar{Z}_{m}}}{\bar{Z}^2_{m}-\bar{Z}^2_{m+l}}
  \\
  &\times\biggl[(i\omega_m+\mu_{d}^{B}-\Sigma_{m}^B)
  (i\omega_m+i\nu_l+\mu_{d}^{A}-\Sigma_{m+l}^A)
  \nonumber\\
  &+(i\omega_m+\mu_{d}^{A}-\Sigma_{m}^A)(i\omega_m+i\nu_l+\mu_{d}^{B}
  -\Sigma_{m+l}^B)\biggr]
  \nonumber\\
  &+2\dfrac{\bar{Z}_{m+l}F_\infty(\bar{Z}_{m+l})
  -\bar{Z}_m F_\infty(\bar{Z}_m)}{\bar{Z}^2_m-\bar{Z}^2_{m+l}}\Biggr\}.
  \nonumber
\end{align}
Then, after substituting this expression for $\Pi_{m,m+l}$ into Eq.~(\ref{Pi}) and replacing the summation over fermionic Matsubara frequencies by integrals over the real frequency axis, the total expression for the nonresonant response function equals
\begin{align}\label{resp_b1g}
  &\chi_{NB_{\rm 1g}}(\Omega)=\dfrac{1}{4\pi^{2}}\int\limits_{-\infty}^{+\infty}d\omega
  \left[f(\omega)-f(\omega+\Omega)\right]
  \\
  &\times\Real\Biggl\{\dfrac{\dfrac{F^{*}_\infty[\bar{Z}(\omega+\Omega)]}{\bar{Z}^{*}
  (\omega+\Omega)}-\dfrac{F_\infty[\bar{Z}(\omega)]}{\bar{Z}(\omega)}}
  {\bar{Z}^{2}(\omega)-[\bar{Z}^{*}(\omega+\Omega)]^{2}}
  \nonumber\\
  &\times
  \biggl([\omega+\mu_{d}^{B}-\Sigma^{B}(\omega)]
  [\omega+\Omega+\mu_{d}^{A}-\Sigma^{A*}(\omega+\Omega)]
  \nonumber \\
  &+[\omega+\mu_{d}^{A}-\Sigma^{A}(\omega)]
  [\omega+\Omega+\mu_{d}^{B}-\Sigma^{B*}(\omega+\Omega)]\biggl)
  \nonumber\\
  &+2\dfrac{\bar{Z}^{*}(\omega+\Omega)F_\infty^{*}[\bar{Z}(\omega+\Omega)]-
  \bar{Z}(\omega)F_\infty[\bar{Z}(\omega)]}
  {\bar{Z}^{2}(\omega)-[\bar{Z}^{*}(\omega+\Omega)]^{2}}
  \nonumber\\
  &-\dfrac{\dfrac{F_\infty[\bar{Z}(\omega+\Omega)]}{\bar{Z}(\omega+\Omega)}
  -\dfrac{F_\infty[\bar{Z}(\omega)]}{\bar{Z}(\omega)}}{\bar{Z}^{2}(\omega)
  -\bar{Z}^{2}(\omega+\Omega)}
  \nonumber\\
  &\times
  \biggl([\omega+\mu_{d}^{B}-\Sigma^{B}(\omega)]
  [\omega+\Omega+\mu_{d}^{A}-\Sigma^{A}(\omega+\Omega)]
  \nonumber \\
  &+[\omega+\mu_{d}^{A}-\Sigma^{A}(\omega)]
  [\omega+\Omega+\mu_{d}^{B}-\Sigma^{B}(\omega+\Omega)]\biggr)\nonumber\\
  &-2\dfrac{\bar{Z}(\omega+\Omega)F_\infty[\bar{Z}(\omega+\Omega)]
  -\bar{Z}(\omega)F_\infty[\bar{Z}(\omega)]}
  {\bar{Z}^{2}(\omega)-\bar{Z}^{2}(\omega+\Omega)}\Biggr\}.
  \nonumber
\end{align}
One can check that this expression for the Raman response function (for $B_{\rm 1g}$ symmetry) in the CDW phase is connected with the one for the optical conductivity\cite{matveev} by the Shastry-Shraiman relation\cite{shastry_shraiman}
\begin{equation}\label{eq: shastry_shraiman}
  \chi_{NB_{\rm 1g}}(\Omega)=\Omega\sigma(\Omega),
\end{equation}
indicating that this relation continues to hold even in the ordered phases. This formula holds for all models, because it does not depend on the vertex function.

The case of $A_{\rm 1g}$ symmetry has both terms of the Feynman diagram of Fig.~\ref{fig: polarization} contributing to the expression for the nonresonant response function. According to the form of the stress-tensor factor, the summation over momentum of the bare loop yields
  \begin{align}\label{a1gP1}
  \Pi&_{m,m+l}^{(1)}=\dfrac{1}{2}\Biggl\{
  \dfrac{\bar{Z}_{m+l}F_{\infty}(\bar{Z}_{m+l})-\bar{Z}_{m}F_{\infty}(\bar{Z}_{m})}
  {\bar{Z}^{2}_{m}-\bar{Z}^{2}_{m+l}}
  \\
  &\times\biggl[(i\omega_m+\mu_{d}^{B}-\Sigma_{m}^B)
  (i\omega_m+i\nu_l+\mu_{d}^{A}-\Sigma_{m+l}^A)
  \nonumber \\
  &+(i\omega_m+\mu_{d}^{A}-\Sigma_{m}^A)(i\omega_m+i\nu_l+\mu_{d}^{B}
  -\Sigma_{m+l}^B)\biggr]
  \nonumber \\
  &+1+\dfrac{\bar{Z}^{3}_{m+l}F_{\infty}(\bar{Z}_{m+l})
  -\bar{Z}^{3}_{m}F_{\infty}(\bar{Z}_{m})}{\bar{Z}^{2}_{m}-\bar{Z}^{2}_{m+l}}\Biggr\}
  \nonumber
  \end{align}
which is different from the one for the $B_{\rm 1g}$ symmetry in Eq.~(\ref{b1gP1}).

\begin{figure}[tb]
   \raisebox{-.45\height}{\includegraphics[scale=0.45]{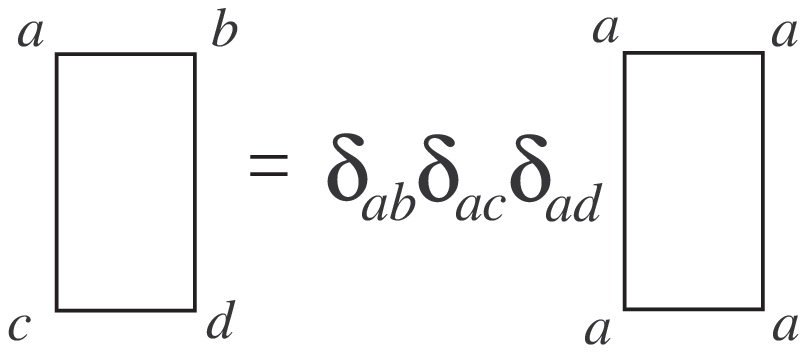}},
   \raisebox{-.45\height}{\includegraphics[scale=0.45]{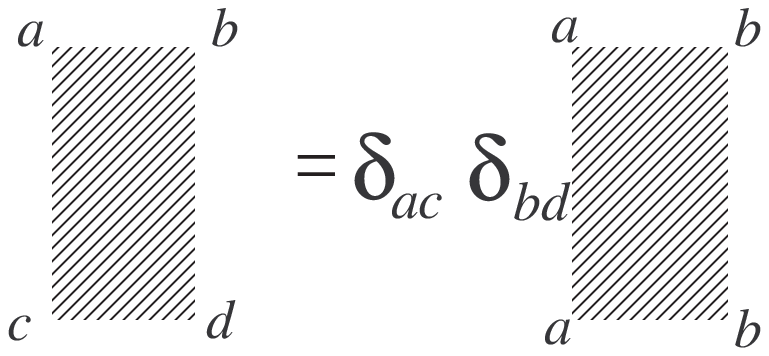}}
   \caption{The irreducible charge vertex becomes local in DMFT and, accordingly,
   the reducible charge vertex depends only on two sublattice indexes.}
   \label{fig: Gamma_delta}
\end{figure}
\begin{figure}[tb]
   ${\tilde{\Gamma}^{ab}}=$
   \raisebox{-.45\height}{\includegraphics[scale=0.45]{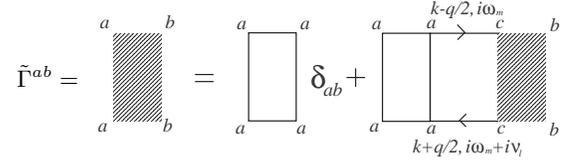}}
   \caption{The Bethe-Salpeter equation for the reducible charge vertex in the CDW
   chessboard phase.}
   \label{fig: Bethe-Salpeter}
   \end{figure}

The second term of the Feynman diagram in Fig.~\ref{fig: response} describes the charge screening effects through the reducible charge vertex, which is defined from the irreducible one. In the DMFT approach, the irreducible charge vertex ${\Gamma}_{a}$ is local and different for different sublattices (see Fig.~\ref{fig: Gamma_delta}); nevertheless, it has the same functional form as in the uniform phase, which is equal to\cite{charge_vertex}
\begin{align}\label{eq: Gamma}
  &\Gamma_{a}(i\omega_{m},i\omega_{m'};i\nu_{l})
  =\delta_{m m'}\Gamma^{a}_{m,m+l}
  \\
  &\Gamma^{a}_{m,m+l}=\dfrac{1}{T}\dfrac{\Sigma^{a}_{m}-\Sigma^{a}_{m+l}}
   {G^{aa}_{m}-G^{aa}_{m+l}}
  \nonumber
\end{align}
for the Falicov-Kimball model (an explicit formula for other models is not known).
This expression also follows from the partially integrated Ward identity, derived by Janis\cite{janis}. Accordingly, the reducible charge vertex in the CDW chessboard phase depends on two sublattice indexes and is defined by the Bethe-Salpeter equation in Fig.~\ref{fig: Bethe-Salpeter}
\begin{equation}\label{eq: Gamma_BS}
  \tilde{\Gamma}^{ab}_{\bm q,m,m+l}=\delta_{ab}\Gamma^{a}_{m,m+l}
  +T\Gamma^{a}_{m,m+l} \sum_c \chi^{ac}_{\bm q,m,m+l}
  \tilde{\Gamma}^{cb}_{\bm q,m,m+l},
\end{equation}
where we introduce the bare susceptibility
\begin{equation}\label{eq: chi_bare}
  \chi^{ab}_{\bm q,m,m+l}= -\frac{1}{N}\sum_{\bm k} G^{ab}_{\bm k, m} G^{ba}_{\bm k+\bm q, m+l}.
\end{equation}

\begin{widetext}
Now, the generalized polarization can be presented in a compact matrix form as follows
\begin{align}\label{eq: Pi_2}
   \Pi_{\bm q,m,m+l}^{(2)}&= \dfrac{1}{N}\sum\limits_{\bm k}
  \left[
   \begin{array}{cc}
   \bar{\gamma}_{\bm k}
   & \bar{\gamma}_{\bm k} \\
   \end{array}
  \right]
  \left\|
   \begin{array}{cc}
   G_{\bm k-\tfrac{\bm q}2,m}^{AA}G_{\bm k+\tfrac{\bm q}2,m+l}^{AB}
   & G_{\bm k-\tfrac{\bm q}2,m}^{AB}G_{\bm k+\tfrac{\bm q}2,m+l}^{BB} \\
   G_{\bm k-\tfrac{\bm q}2,m}^{BA}G_{\bm k+\tfrac{\bm q}2,m+l}^{AA}
   & G_{\bm k-\tfrac{\bm q}2, m}^{BB}G_{\bm k+\tfrac{\bm q}2,m+l}^{BA} \\
   \end{array}
  \right\|
  \\
  &\times T\left\|
   \begin{array}{cc}
   \tilde{\Gamma}^{AA}_{\bm q,m,m+l} & \tilde{\Gamma}^{AB}_{\bm q,m,m+l} \\
   \tilde{\Gamma}^{BA}_{\bm q,m,m+l} & \tilde{\Gamma}^{BB}_{\bm q,m,m+l} \\
   \end{array}
  \right\|
\dfrac{1}{N}\sum\limits_{\bm k'}\left\|
   \begin{array}{cc}
   G_{\bm k'-\tfrac{\bm q}2,m}^{AA}G_{\bm k'+\tfrac{\bm q}2,m+l}^{BA}
   & G_{\bm k'-\tfrac{\bm q}2,m}^{AB}G_{\bm k'+\tfrac{\bm q}2,m+l}^{AA} \\
   G_{\bm k'-\tfrac{\bm q}2,m}^{BA}G_{\bm k'+\tfrac{\bm q}2,m+l}^{BB}
   & G_{\bm k'-\tfrac{\bm q}2,m}^{BB}G_{\bm k'+\tfrac{\bm q}2,m+l}^{AB} \\
   \end{array}
  \right\|\left[
   \begin{array}{c} \bar{\gamma}_{\bm k'} \\
   \bar{\gamma}_{\bm k'}
   \end{array}
  \right].
  \nonumber
\end{align}
The next step is to put $\bm q=0$, expand the expression via partial fractions with respect to the band energy $\epsilon_{\bm k}$ and calculate the sums over momentum $\bm k$. After some tedious algebra, we obtain the final expression for $\Pi_{m,m+l}^{(2)}$
\begin{align}
  &\Pi_{m,m+l}^{(2)}
  =\dfrac{1}{\Delta_{m,m+l}}\,\dfrac{\left[\bar{Z}_{m+l}F_{\infty}(\bar{Z}_{m+l})
  -\bar{Z}_{m}F_{\infty}(\bar{Z}_{m})\right]^{2}}
  {\bar{Z}^{2}_{m}-\bar{Z}^{2}_{m+l}}
  \\
  &\times \Biggl\{\left[i(2\omega_m+\nu_l)+2\mu_{d}^{B}-\Sigma_{m}^B-\Sigma_{m+l}^B\right]
  \dfrac{\Sigma_{m}^A-\Sigma_{m+l}^A}{G_{m}^{AA}-G_{m+l}^{AA}}
  \Biggl(\left[i(2\omega_m+\nu_l)+2\mu_{d}^{B}-\Sigma_{m}^B-\Sigma_{m+l}^B\right]
  \nonumber \\
  &\times\left\{\left[\bar{Z}^{2}_{m}-\bar{Z}^{2}_{m+l}\right]
  -(i\omega_m+\mu_{d}^{B}-\Sigma_{m}^B)
  (i\omega_m+i\nu_l+\mu_{d}^{B}-\Sigma_{m+l}^{B})
  \left[\dfrac{F_{\infty}(\bar{Z}_{m+l})}{\bar{Z}_{m+l}}
  -\dfrac{F_{\infty}(\bar{Z}_{m})}{\bar{Z}_{m}}\right]
  \dfrac{\Sigma_{m}^B-\Sigma_{m+l}^B}{G_{m}^{BB}-G_{m+l}^{BB}}\right\}
  \nonumber \\
  &+\left[i(2\omega_m+\nu_l)+2\mu_{d}^{A}-\Sigma_{m}^A-\Sigma_{m+l}^A\right]
  \dfrac{\Sigma_{m}^A-\Sigma_{m+l}^A}{G_{m}^{AA}-G_{m+l}^{AA}}
  \left[\bar{Z}_{m+l}F_{\infty}(\bar{Z}_{m+l})
  -\bar{Z}_{m}F_{\infty}(\bar{Z}_{m})\right]\Biggr)
  \nonumber \\
  &\left[i(2\omega_m+\nu_l)+2\mu_{d}^{A}-\Sigma_{m}^A-\Sigma_{m+l}^A\right]
  \dfrac{\Sigma_{m}^B-\Sigma_{m+l}^B}{G_{m}^{BB}-G_{m+l}^{BB}}
  \Biggl(\left[i(2\omega_m+\nu_l)+2\mu_{d}^{A}-\Sigma_{m}^A-\Sigma_{m+l}^A\right]
  \nonumber \\
  &\times\left\{\left[\bar{Z}^{2}_{m}-\bar{Z}^{2}_{m+l}\right]
  -(i\omega_m+\mu_{d}^{A}-\Sigma_{m}^A)
  (i\omega_m+i\nu_l+\mu_{d}^{A}-\Sigma_{m+l}^{A})
  \left[\dfrac{F_{\infty}(\bar{Z}_{m+l})}{\bar{Z}_{m+l}}
  -\dfrac{F_{\infty}(\bar{Z}_{m})}{\bar{Z}_{m}}\right]
  \dfrac{\Sigma_{m}^A-\Sigma_{m+l}^A}{G_{m}^{AA}-G_{m+l}^{AA}}\right\}
  \nonumber \\
  &+\left[i(2\omega_m+\nu_l)+2\mu_{d}^{B}-\Sigma_{m}^B-\Sigma_{m+l}^B\right]
  \dfrac{\Sigma_{m}^B-\Sigma_{m+l}^B}{G_{m}^{BB}-G_{m+l}^{BB}}
  \left[\bar{Z}_{m+l} F_{\infty}(\bar{Z}_{m+l})
  -\bar{Z}_{m} F_{\infty}(\bar{Z}_{m})\right]\Biggr)
  \Biggr\},
  \nonumber
\end{align}
\end{widetext}
where
\begin{equation}
  \Delta_{m,m+l}=\det\left\|\delta_{ab}-T\Gamma^{a}_{m,m+l}\chi^{ab}_{\bm q=0,m,m+l}\right\|
\end{equation}
comes from the solution of the Bethe-Salpeter equation in Eq.~(\ref{eq: Gamma_BS}). Finally, the total expression for the generalized polarization is obtained as the sum of the two contributions
\begin{equation}
\Pi_{m,m+l}=\Pi_{m,m+l}^{(1)}+\Pi_{m,m+l}^{(2)}.
\end{equation}
Next, we perform an analytical continuation (which is straightforward because we have the appropriate functional forms which allow us to replace Matsubara frequencies by real frequencies) and substitute into Eq.~(\ref{eq: resp_nonres}) which yields the final expression for the nonresonant Raman response function in the $A_{\rm 1g}$ channel.  This step is completely straightforward, so we do not write down the final expressions in terms of integrals over the real frequency.

For other many-body models, such as the Hubbard, or Holstein models, the analysis is more complicated because the vertex is not diagonal, and the analytic continuation will not be possible on a formal level, if the charge vertex cannot be expressed as functions of the Green's function and self-energy.  Nevertheless, one can perform the analytic continuation numerically, and thereby solve the problem, or one can approximate the vertex using different analytic approximations (such as those from perturbation theory) and then formally complete the analytic continuation.

\subsection{X-ray scattering: $\bm q\ne0$}

In the case of inelastic X-ray scattering, the incident photon exchanges both energy and momentum with the electronic matter. The entire formalism derived for Raman scattering remains the same as described above and there is no need to rewrite it for this case. The only difference is in the summations over momentum. The Feynman diagrams in Fig.~\ref{fig: response} together with the Bethe-Salpeter equation in Fig.~\ref{fig: Bethe-Salpeter} contain several  momentum summations which can be evaluated separately\cite{deveraux_freericks_mccormack}. First, the bare susceptibility in Eq.~(\ref{eq: chi_bare}), which enters the Bethe-Salpeter equation for the total charge vertex in Eq.~(\ref{eq: Gamma_BS}), contains the following components
\begin{align}
  &\chi_{\bm q,m,m+l}^{AA}=\frac{(i\omega_m+\mu_d^B-\Sigma_m^B)(i\omega_{m+l}+\mu_d^B-\Sigma_{m+l}^B)}
   {2\bar{Z}_m\bar{Z}_{m+l}}
  \\
  &\times\left[\chi_{0}(\bar{Z}_{m},\bar{Z}_{m+l},\bm q)
     -\chi_{0}(\bar{Z}_{m},-\bar{Z}_{m+l},\bm q)\right],
  \nonumber
\end{align}
\begin{align}
  &\chi_{\bm q,m,m+l}^{BB}=\frac{(i\omega_m+\mu_d^A-\Sigma_m^A)(i\omega_{m+l}+\mu_d^A-\Sigma_{m+l}^A)}
   {2\bar{Z}_m\bar{Z}_{m+l}}
  \\
  &\times\left[\chi_{0}(\bar{Z}_{m},\bar{Z}_{m+l},\bm q)
     -\chi_{0}(\bar{Z}_{m},-\bar{Z}_{m+l},\bm q)\right],
  \nonumber
\end{align}
and
\begin{align}
  &\chi_{\bm q,m,m+l}^{AB}=\chi_{\bm q,m,m+l}^{BA}
  \\
  &=\frac{1}{2}\left[\chi_{0}(\bar{Z}_{m},\bar{Z}_{m+l},\bm q)
     +\chi_{0}(\bar{Z}_{m},-\bar{Z}_{m+l},\bm q)\right],
  \nonumber
\end{align}
where
\begin{align}
  &\chi_{0}(\bar{Z}_{m},\bar{Z}_{m+l},\bm q)
   =\chi_{0}(-\bar{Z}_{m},-\bar{Z}_{m+l},\bm q)
  \\
  &=-\dfrac{1}{N}\sum_{k}\dfrac{1}{\bar{Z}_{m}-\epsilon_{\bm k-\tfrac{\bm q}2}}
  \dfrac{1}{\bar{Z}_{m+l}-\epsilon_{\bm k+\tfrac{\bm q}2}}
  \nonumber \\
  &=-\dfrac{1}{\sqrt{1-X^{2}}}
  \int\limits_{-\infty}^{+\infty}
  \dfrac{d\epsilon}{\bar{Z}_{m+l}-\epsilon}\,\rho(\epsilon)
  F_\infty\left(\dfrac{\bar{Z}_{m}-\epsilon X}{\sqrt{1-X^{2}}}\right).
  \nonumber
\end{align}
Here, the function $F_\infty(Z)$ is the Hilbert transform of the hypercubic density of states as defined in Eq.~(\ref{eq: Hilb_trans}) and all the transferred momentum dependence is only through the quantity
\begin{equation}\label{eq: X}
  X=\dfrac{1}{D}\sum^{D}\limits_{p=1}\cos q_{p}.
\end{equation}

The second diagram in Fig.~\ref{fig: response} contains summations over $\bm k$ and $\bm k^\prime$ which involve stress-tensor amplitudes $\bar{\gamma}_{\bm k}$
\begin{align}
  \chi_{\bm q,m,m+l}^{a}&=\frac1N\sum_{\bm k}\bar{\gamma}_{\bm k}
  \\
  &\times\left[G_{\bm k-\tfrac{\bm q}2,m}^{Aa}G_{\bm k+\tfrac{\bm q}2,m+l}^{aB}
  +G_{\bm k-\tfrac{\bm q}2,m}^{Ba}G_{\bm k+\tfrac{\bm q}2,m+l}^{aA}\right]
  \nonumber
\end{align}
and there are two different terms:
\begin{align}
  \chi_{\bm q,m,m+l}^{A}&=\frac{i\omega_m+\mu_d^B-\Sigma_m^B}{2\bar{Z}_m}
  \\
  &\times\left[\chi_{0}^{\prime}(\bar{Z}_{m},\bar{Z}_{m+l},\bm q)
   +\chi_{0}^{\prime}(\bar{Z}_{m},-\bar{Z}_{m+l},\bm q)\right]
   \nonumber\\
  &+\frac{i\omega_{m+l}+\mu_d^B-\Sigma_{m+l}^B}{2\bar{Z}_{m+l}}
  \nonumber\\
  &\times\left[\chi_{0}^{\prime}(\bar{Z}_{m},\bar{Z}_{m+l},\bm q)
     -\chi_{0}^{\prime}(\bar{Z}_{m},-\bar{Z}_{m+l},\bm q)\right]
  \nonumber
\end{align}
and
\begin{align}
  \chi_{\bm q,m,m+l}^{B}&=\frac{i\omega_m+\mu_d^A-\Sigma_m^A}{2\bar{Z}_m}
   \\
  &\times\left[\chi_{0}^{\prime}(\bar{Z}_{m},\bar{Z}_{m+l},\bm q)
   +\chi_{0}^{\prime}(\bar{Z}_{m},-\bar{Z}_{m+l},\bm q)\right]
   \nonumber\\
  &+\frac{i\omega_{m+l}+\mu_d^A-\Sigma_{m+l}^A}{2\bar{Z}_{m+l}}
   \nonumber\\
  &\times\left[\chi_{0}^{\prime}(\bar{Z}_{m},\bar{Z}_{m+l},\bm q)
     -\chi_{0}^{\prime}(\bar{Z}_{m},-\bar{Z}_{m+l},\bm q)\right],
  \nonumber
\end{align}
where
\begin{align}
  &\chi^\prime_{0}(\bar{Z}_{m},\bar{Z}_{m+l},\bm q)
  =-\chi^\prime_{0}(-\bar{Z}_{m},-\bar{Z}_{m+l},\bm q)
  \\
  &=-\dfrac{1}{N}\sum_{\bm k}\bar{\gamma}_{\bm k}\dfrac{1}
  {\bar{Z}_{m}-\epsilon_{\bm k-\tfrac{\bm q}2}}
  \dfrac{1}{\bar{Z}_{m+l}-\epsilon_{\bm k+\tfrac{\bm q}2}}
  \nonumber\\
  &=\dfrac{X^\prime}{1+X}\Bigl\{\left[\bar{Z}_{m}+\bar{Z}_{m+l}\right]
  \chi_{0}(\bar{Z}_{m},\bar{Z}_{m+l},\bm q)
  \nonumber \\
  &+F_{\infty}[\bar{Z}(i\omega_{m})]+F_{\infty}[\bar{Z}(i\omega_{m+l})]\Bigr\}.
  \nonumber
\end{align}
Here the new momentum dependent quantity $X^\prime$ is
\begin{equation}
  X^\prime=\dfrac{1}{D}\sum^{D}\limits_{p=1}\alpha_{p}\cos\dfrac{q_{p}}{2}
\end{equation}
with $\alpha_{p}=1$ for $A_{\rm 1g}$ symmetry and $\alpha_{p}=(-1)^{p}$ for $B_{\rm 1g}$ symmetry.
Now we can find exact expression for the vertex corrections defined by Eq.~(\ref{eq: Pi_2}) in the following form
\begin{align}\label{eq: Pi2_NIXS}
  &\Pi^{(2)}_{\bm q,m,m+l}=\frac{1}{\Delta_{\bm q,m,m+l}}
  \\
  &\times\Bigl[
  \chi^A_{\bm q,m,m+l}T\Gamma^A_{m,m+l}\chi^{AB}_{\bm q,m,m+l}T\Gamma^B_{m,m+l}\chi^{B}_{\bm q,m,m+l}
  \nonumber\\
  &+\chi^A_{\bm q,m,m+l}\left(1-T\Gamma^B_{m,m+l}\chi^{BB}_{\bm q,m,m+l}\right) T\Gamma^A_{m,m+l}\chi^{A}_{\bm q,m,m+l}
  \nonumber\\
  &+\chi^B_{\bm q,m,m+l}\left(1-T\Gamma^A_{m,m+l}\chi^{AA}_{\bm q,m,m+l}\right) T\Gamma^B_{m,m+l}\chi^{B}_{\bm q,m,m+l}
  \nonumber\\
  &+\chi^B_{\bm q,m,m+l}T\Gamma^B_{m,m+l}\chi^{BA}_{\bm q,m,m+l}T\Gamma^A_{m,m+l}\chi^{A}_{\bm q,m,m+l}
  \Bigr],
  \nonumber
\end{align}
where
\begin{align}
  &\Delta_{\bm q,m,m+l}
  \\
  &=\left(1-T\Gamma^A_{m,m+l}\chi^{AA}_{\bm q,m,m+l}\right)
   \left(1-T\Gamma^B_{m,m+l}\chi^{BB}_{\bm q,m,m+l}\right)
  \nonumber\\
  &-T\Gamma^A_{m,m+l}\chi^{AB}_{\bm q,m,m+l}T\Gamma^B_{m,m+l}\chi^{BA}_{\bm q,m,m+l}
  \nonumber
\end{align}

Finally, the bare loop contribution of the first diagram in Fig.~\ref{fig: response} contains summations over momentum $\bm k$ of the product of two Green functions and the square of the stress-tensor factor.
It is equal to
\begin{align}
  &\Pi^{(1)}_{\bm q,m,m+l}=\bar{\chi}_{0}(\bar{Z}_{m},\bar{Z}_{m+l},\bm q)
  +\bar{\chi}_{0}(\bar{Z}_{m},-\bar{Z}_{m+l},\bm q)
  \label{eq: Pi1X}\\
  &+\frac{1}{2\bar{Z}_{m}\bar{Z}_{m+l}}
  \Bigl[(i\omega_m+\mu_d^A-\Sigma_m^A)
   (i\omega_{m+l}  +\mu_d^A-\Sigma_{m+l}^A)
  \nonumber\\
  &+(i\omega_m+\mu_d^B-\Sigma_m^B)(i\omega_{m+l}+\mu_d^B-\Sigma_{m+l}^B)\Bigr]
  \nonumber\\
  &\times\left[\bar{\chi}_{0}(\bar{Z}_{m},\bar{Z}_{m+l},\bm q)
  -\bar{\chi}_{0}(\bar{Z}_{m},-\bar{Z}_{m+l},\bm q)\right]
  \nonumber
\end{align}
 and expressed in terms of $\chi_{0}$ as follows
\begin{align}
  &\bar{\chi}_{0}(\bar{Z}_{m},\bar{Z}_{m+l},\bm q)
  =\bar{\chi}_{0}(-\bar{Z}_{m},-\bar{Z}_{m+l},\bm q)
  \label{eq: chi0bar}\\
  &=-\dfrac{1}{N}\sum_{\bm k}\bar{\gamma}^{2}_{\bm k}\dfrac{1}
  {\bar{Z}_{m}-\epsilon_{\bm k-\tfrac{\bm q}2}}
  \dfrac{1}{\bar{Z}_{m+l}-\epsilon_{\bm k+\tfrac{\bm q}2}}
  \nonumber \\
  &=\chi_{0}(\bar{Z}_{m},\bar{Z}_{m+l},\bm q)
  \left\{\dfrac{t^{*2}}{2}-\dfrac{t^{*2}X^{\prime2}}{1+X}
  +\dfrac{\left[\bar{Z}_{m}+\bar{Z}_{m+l}\right]^{2}X^{\prime2}}
  {(1+X)^{2}}\right\}
  \nonumber \\
  &+\dfrac{X^{\prime2}}{(1+X)^{2}}\left[\bar{Z}_{m}+\bar{Z}_{m+l}\right]
  \left\{F_{\infty}[\bar{Z}(i\omega_{m})]+F_{\infty}[\bar{Z}(i\omega_{m+l})]\right\}
  \nonumber \\
  &+\dfrac{X^{\prime2}}{1+X}\left\{\bar{Z}_{m}F_{\infty}[\bar{Z}(i\omega_{m})]
  +\bar{Z}_{m+l}F_{\infty}[\bar{Z}(i\omega_{m+l})]-2\right\}.
  \nonumber
\end{align}
The expressions for $\chi^\prime_{0}$ and $\bar{\chi}_{0}$ derived above appear to be different from the ones given in Ref.~\onlinecite{deveraux_freericks_mccormack}. In fact, they are identical (but require some significant algebra to show this); the forms presented above are more convenient for numerical calculations.

In contrast to $B_{\rm 1g}$ Raman scattering at $\bm q=0$ which is determined only by the bare loop contributions (Fig.~\ref{fig: polarization}), in the case of inelastic X-ray scattering, we have both terms contributing for all symmetry channels. The different symmetry channels are distinguished only by the different $X^\prime$ factors, and, as a result, different $\chi^\prime_{0}$ and $\bar{\chi}_{0}$ functions. All further numerical calculations are performed by exploiting these three quantities, but the total scheme remains the same. As a result, the generalized polarization in Eq.~(\ref{eq: Pi2_NIXS}) is described in terms of the $\chi_{0}$, $\chi^\prime_{0}$ and $\bar{\chi}_{0}$ functions and applying further analytic continuation to the real axis one can obtain the nonresonant inelastic X-ray scattering response functions. The final expressions are too long to be presented here.

\subsection{Nonresonant inelastic X-ray scattering sum rule}

The sum rule for the nonresonant inelastic scattering response function is as follows:\cite{sum_rule,sum_rule_HM}
\begin{equation}\label{sum_rule_def}
  I=\int_{0}^{+\infty} d\Omega \;\Omega\; \chi_N(\Omega) =
    \frac{\pi}{2}\left\langle\left[\tilde\gamma^{\dag}(\bm q)
    \left[H,\tilde\gamma(\bm q)\right]\right]\right\rangle,
\end{equation}
where for the case of CDW ordering
\begin{align}
  \tilde\gamma(\bm q)&=\sum_{ab}\sum_{ij} t_{ij}^{ab} e^{i\bm Q(\bm R_i^a-\bm R_j^b)}
  e^{-i\frac{\bm q}2 (\bm R_i^a+\bm R_j^b)} \hat{d}_{ia}^{\dag} \hat{d}_{jb},
  \\
  \tilde\gamma^{\dag}(\bm q)&=\sum_{ab}\sum_{ij} t_{ji}^{ba} e^{i\bm Q(\bm R_j^b-\bm R_i^a)}
  e^{i\frac{\bm q}2 (\bm R_i^a+\bm R_j^b)} \hat{d}_{jb}^{\dag} \hat{d}_{ia},
  \nonumber
\end{align}
and the momentum $\bm Q$ determines the symmetry channels
\begin{equation}
  \bm Q = \left\{\begin{array}{ll}
    0 & \text{ for } A_{1g} \\
    (\pi,0,\pi,0,\ldots) & \text{ for } B_{1g}
\end{array}\right..
\end{equation}

After calculating all required commutators, taking the large dimensional limit, and performing some cumbersome transformations (see the Appendix), we obtain a sum rule (first moment of the nonresonant inelastic X-ray scattering response function) which contains two contributions
\begin{equation}
  I = I_K + I_{\Pi}.
\end{equation}
The first contribution comes from the kinetic energy term
\begin{align}\label{sum_rule_kinetic}
  &I_K = 2(1-X)\!\!\int_{-\infty}^{+\infty}\!\! d\omega f(\omega)\Img\left\{
  \frac{t^{*2}}{2}\left[ \bar Z(\omega) F_\infty[\bar{Z}(\omega)] -1 \right] \right.
  \\
  &\left.- X^{\prime2}\left(\left[ \frac{3t^{*2}}2  -
  \bar Z^2(\omega)\right] \left[ \bar Z(\omega) F_\infty[\bar{Z}(\omega)] -1 \right] + \frac{t^{*2}}2 \right)
  \right\}
  \nonumber
\end{align}
and is similar to the one in the uniform case.\cite{sum_rule}
The other one originates from the potential energy term and satisfies
\begin{align}\label{sum_rule_potential}
  I_{\Pi} &= \int_{-\infty}^{+\infty} d\omega f(\omega)\Img\Biggl\{ \sum_a
  \left[ \Sigma^a(\omega) -U n_f^a \right]
\\
  &\times
  \left[
  \frac{t^{*2}}{2}(1-X^{\prime2})G^{aa}(\omega) +X^{\prime2}\lambda^a(\omega) \right]
  \nonumber\\
  &+ X^{\prime2} \left[ \bar Z(\omega) F_\infty[\bar{Z}(\omega)] -1 \right]
  \left[ \Sigma^A(\omega) - \Sigma^B(\omega) \right]^2 \Biggr\}
  \nonumber\\
  &-\frac{\pi}{2} Ut^{*2}(1-X^{\prime2})(n_f^A-n_f^B)(n_d^A-n_d^B).
\nonumber
\end{align}
(This last term is model dependent and would be different for the Hubbard or Holstein model.  We do not provide those formulas here.)
The first contribution in braces has the same shape as the potential energy contribution of the sum rule in the uniform phase.\cite{sum_rule} The other terms appear only in the CDW phase and are proportional to the square of the CDW order parameter $\left(\Delta n_f\right)^2$.

By examining different points in the Brillouin zone (BZ), one can extract information regarding the potential and kinetic-energy contributions or of the order parameter. For instance, in the case of Raman scattering ($\bm q=0$, $X=1$) we have contributions only from the potential-energy term ($I_K=0$), which are different for the $A_{1g}$ symmetry ($X'=1$)
\begin{align}
  I_{\Pi} &= \int_{-\infty}^{+\infty} d\omega f(\omega)\Img\Biggl\{ \sum_a
  \left[ \Sigma^a(\omega) -Un_f^a \right] \lambda^a(\omega)
  \\
  &+ \left[ \bar Z(\omega) F_\infty[\bar{Z}(\omega)] -1 \right]
  \left[ \Sigma^A(\omega) - \Sigma^B(\omega) \right]^2 \Biggr\},
\nonumber
\end{align}
and for the $B_{1g}$ symmetry ($X'=0$)
\begin{align}
  I_{\Pi} &= \int_{-\infty}^{+\infty} d\omega f(\omega)\Img\Biggl\{ \sum_a
  \left[ \Sigma^a(\omega) -Un_f^a \right] \frac{t^{*2}}{2}G^{aa}(\omega) \Biggr\}
  \\
  &-\frac{\pi}{2} Ut^{*2}(n_f^A-n_f^B)(n_d^A-n_d^B).
\nonumber
\end{align}
For other points in the BZ (inelastic X-ray scattering), we have contributions from both the kinetic and potential-energy terms. For instance, for the case of $B_{1g}$ symmetry along the BZ-diagonal [$\bm q=(q,q,q,q,\ldots)$, $-1\le X \le 1$, $X'=0$] and for all symmetry channels at the BZ corner [$\bm q=(\frac{\pi}{a},\frac{\pi}{a},\frac{\pi}{a},\frac{\pi}{a},\ldots)$, $X=-1$, $X'=0$] we have:
\begin{align}
  I_K &= 2(1-X)\int_{-\infty}^{+\infty} d\omega f(\omega)\Img\left\{
  \frac{t^{*2}}{2}\left[ \bar Z(\omega) F_\infty[\bar{Z}(\omega)] -1 \right]
  \right\},
  \\
  I_{\Pi} &= \int_{-\infty}^{+\infty} d\omega f(\omega)\Img\Biggl\{ \sum_a
  \left[ \Sigma^a(\omega) -Un_f^a \right]
  \frac{t^{*2}}{2}G^{aa}(\omega) \Biggr\}
  \\
  &-\frac{\pi}{2} Ut^{*2}(n_f^A-n_f^B)(n_d^A-n_d^B).
  \nonumber
\end{align}
One can see, that in this case the kinetic-energy contribution is equal (up to an overall constant) to the average kinetic energy which also enters the sum rule for optical conductivity.\cite{matveev}

\section{Numerical results}

\begin{figure}[htb]
   \includegraphics[scale=0.7]{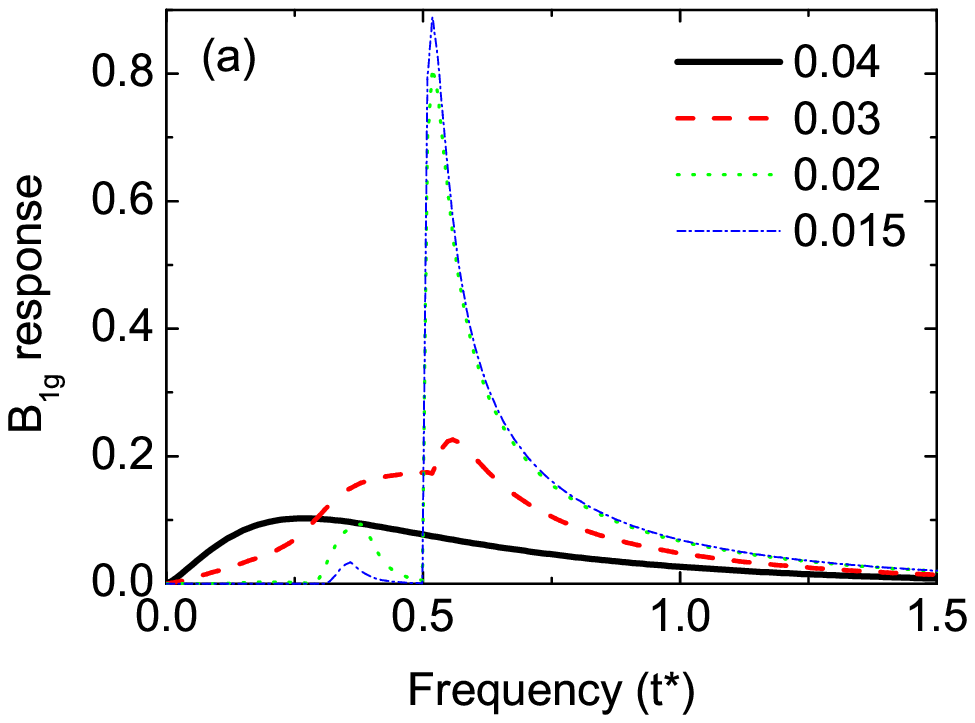}\\
   \includegraphics[scale=0.7]{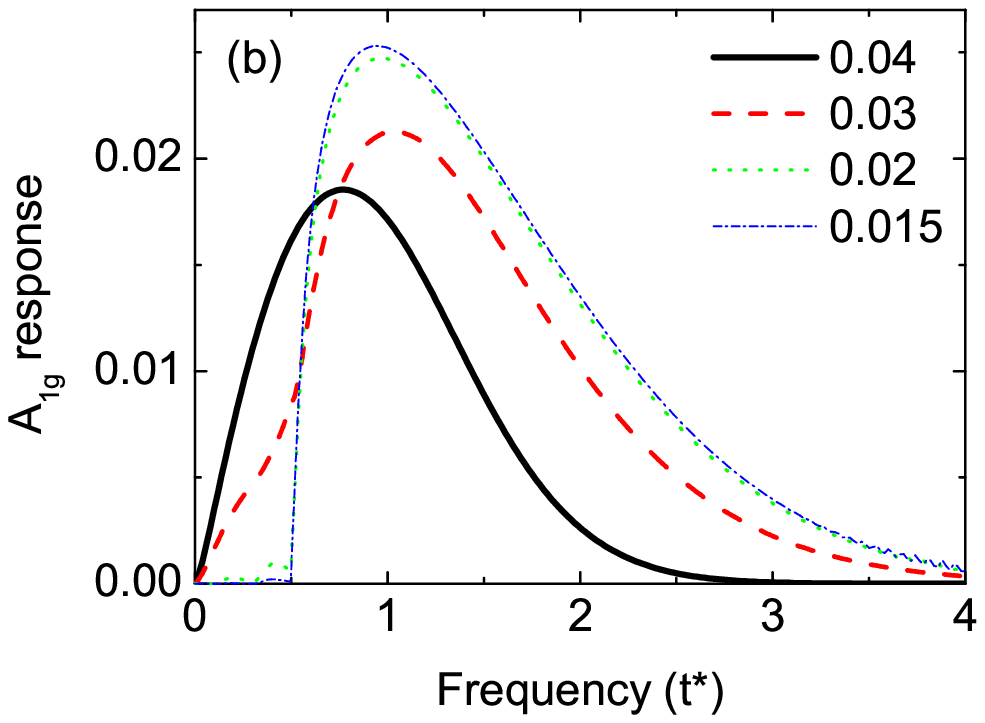}
   \caption{(Color online) Nonresonant Raman response function for the two symmetry channels [(a) being the $B_{\rm 1g}$ channel and (b) being the $A_{\rm 1g}$ channel] in a dirty metal with  $U=0.5$. The set of curves corresponds to a range of temperatures from the uniform to the ordered phase ($T_{c}\approx 0.034$).}
   \label{fig: u05resp}
\end{figure}

\begin{figure}[htb]
   \includegraphics[scale=0.7]{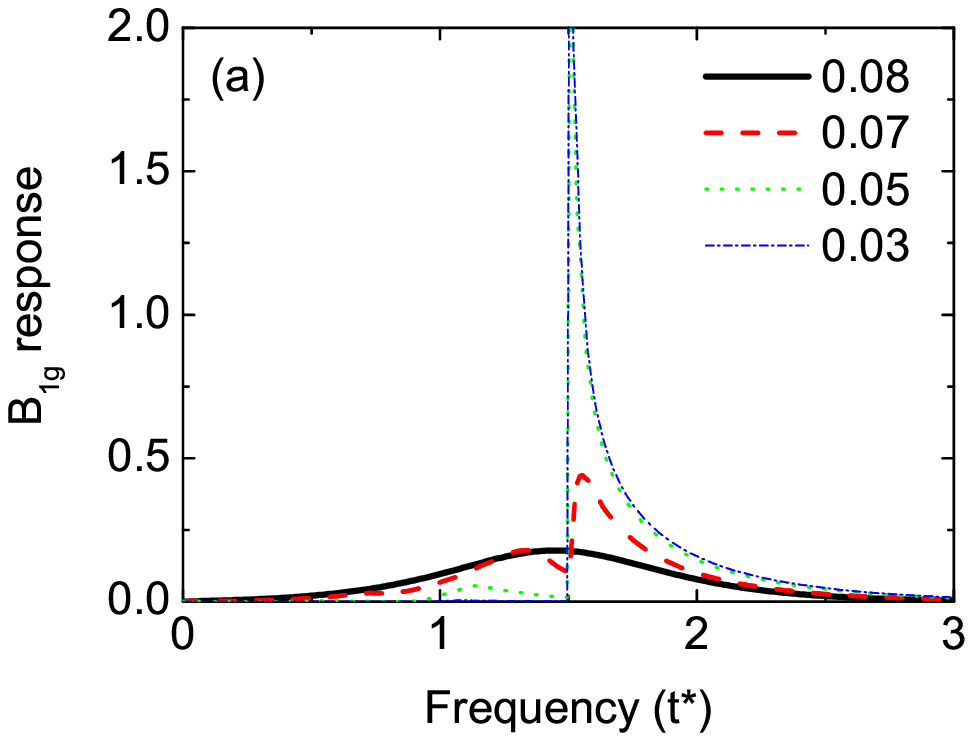}\\
   \includegraphics[scale=0.7]{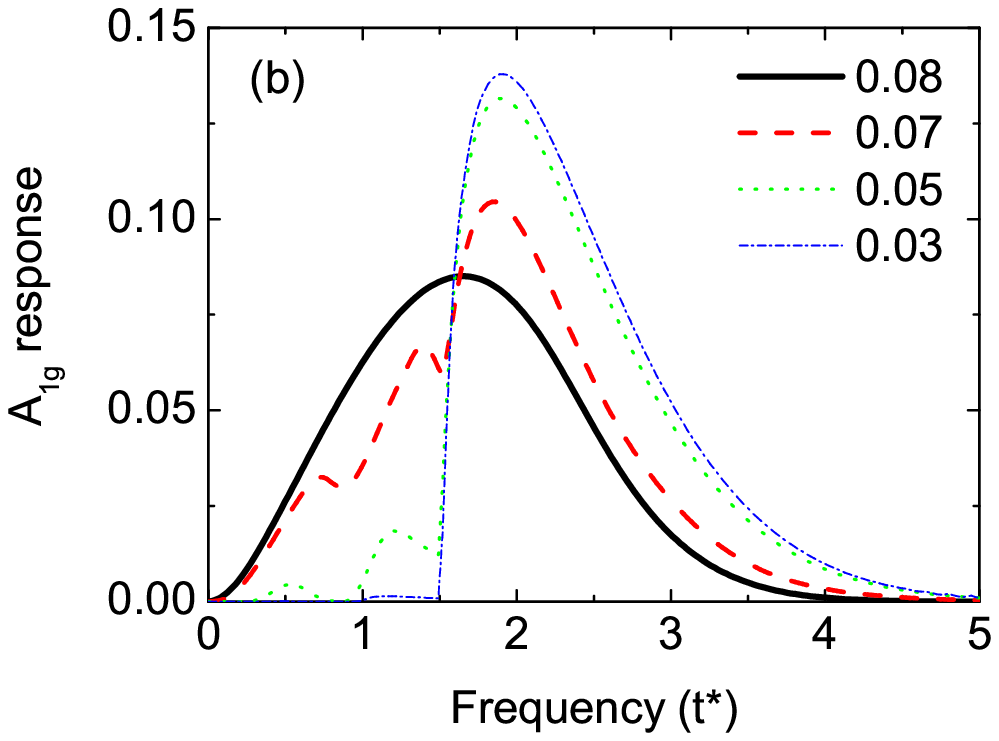}
   \caption{(Color online) Nonresonant Raman response function for the two symmetry channels in a near critical Mott insulator with  $U=1.5$. The set of curves corresponds to a range of temperatures from the uniform to the ordered phase ($T_{c}\approx 0.075$).}
   \label{fig: u15resp}
\end{figure}

\begin{figure}[htb]
   \includegraphics[scale=0.7]{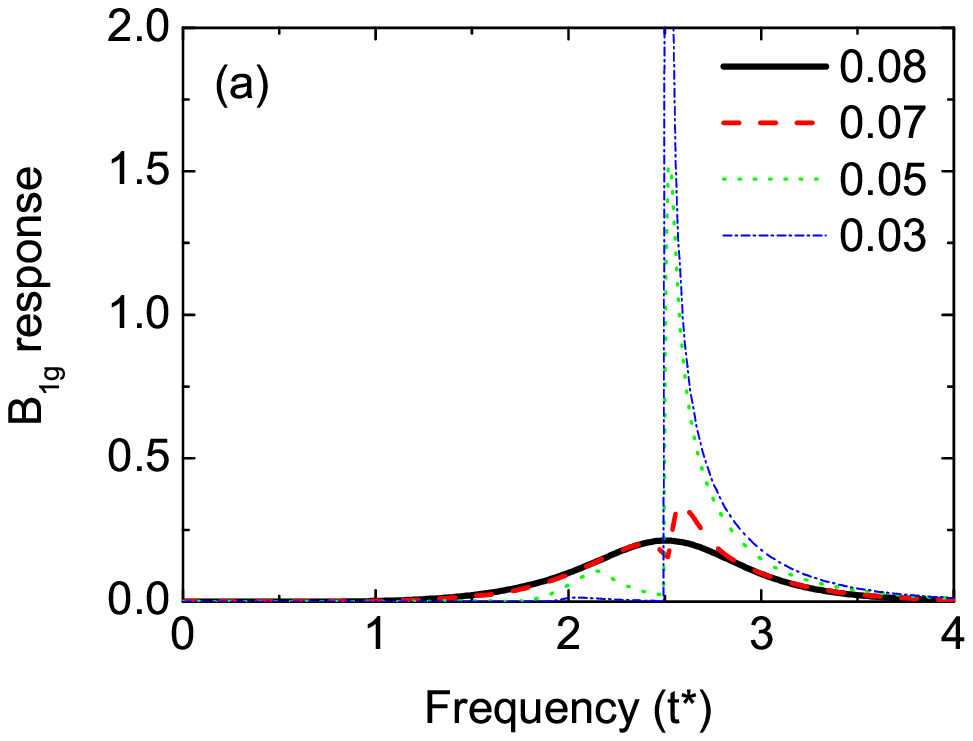}\\
   \includegraphics[scale=0.7]{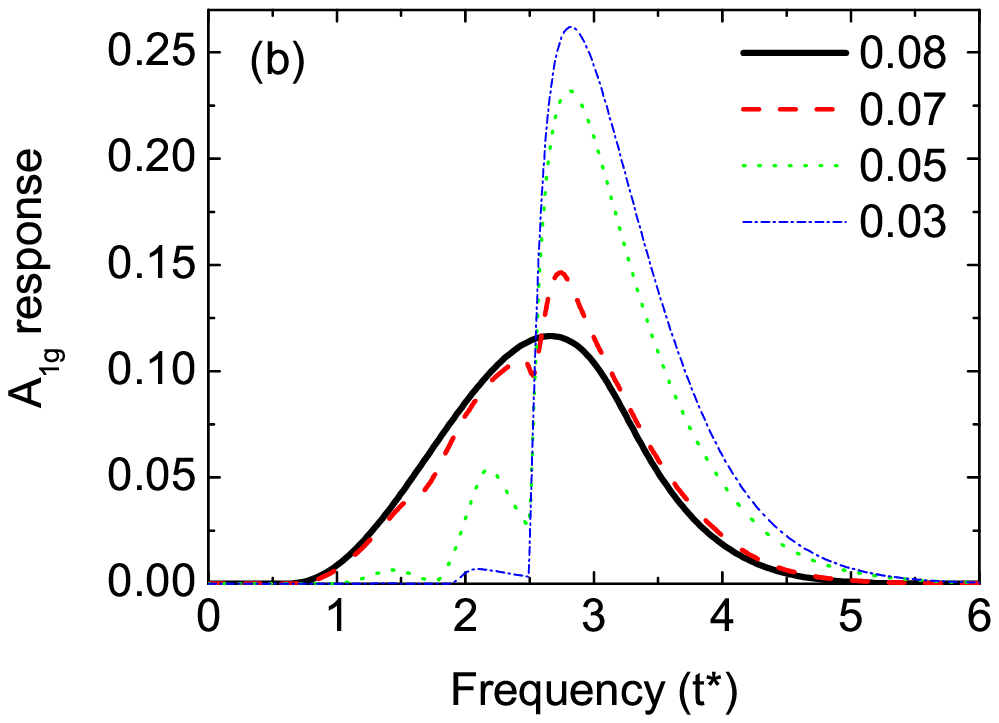}
   \caption{(Color online) Nonresonant Raman response function for the two symmetry channels in a moderate gap Mott insulator with  $U=2.5$. The set of curves corresponds to a range of temperatures from the uniform to the ordered phase ($T_{c}\approx 0.072$).}
   \label{fig: u25resp}
\end{figure}

We begin with an analysis of nonresonant Raman scattering in the CDW phase. We present results for the cases of a dirty metal with $U=0.5$ (Fig.~{\ref{fig: u05resp}}), a near critical Mott insulator with $U=1.5$ (Fig.~{\ref{fig: u15resp}}), and a moderate gap Mott insulator with $U=2.5$ (Fig.~\ref{fig: u25resp}).

In Fig.~{\ref{fig: u05resp}}, we plot the Raman response function for different temperatures in the case of a dirty metal with $U=0.5$. At temperatures higher then the critical one for CDW order, we see the expected behavior for a dirty metal: namely, there is a peak at low energy and a spread on the order of the metallic bandwidth. The system does not have a low energy Fermi-liquid peak, because it is not a Fermi liquid. Below the critical temperature, when the CDW gap arises, the shape of the response function changes significantly. The main peak is shifted to higher frequency at $\Omega\approx U$, which corresponds to transitions between the lowest band at $\omega\le -U/2$ and the upper band at $\omega\ge U/2$ (see the DOS in Ref.~\onlinecite{matveev}). Two additional peaks at lower frequencies correspond to the transitions from the upper and lower bands onto the subgap states and between the subgap states (which are present for a wide range of temperatures below $T_c$ but above $T=0$). Because the subgap DOS vanishes at $T=0$, these peaks must vanish with $T\to 0$. In addition, because the self-energy becomes a frequency independent constant on each sublattice at $T=0$ ($0$ on one sublattice and $U$ on the other), the irreducible charge vertex, and hence the vertex corrections, also vanish at low temperature. In panel (a), we plot the nonresonant response function for the $B_{\rm 1g}$ symmetry. In this symmetry channel, there is a sharp main peak with a square root singularity at $T=0$. This behavior was already seen in the optical conductivity\cite{matveev}, and follows for the Raman scattering directly from the Shastry-Shraiman relation in Eq.~(\ref{eq: shastry_shraiman}). For the $A_{1g}$ channel, as plotted in panel (b), the response is much smaller and smooth (without sharp singularities) and there are two reasons for this. At high temperatures, in the uniform phase and just below $T_c$, we have the effects of dynamical charge screening for the $A_{1g}$ scattering channel which suppresses the total response. On the other hand, the charge vertex is proportional to $U^2 n_f^a(1-n_f^a)$ (see Ref.~\onlinecite{charge_vertex}) and in the charge-ordered phase, where $n_f^A\to1$ and $n_f^B\to0$ for $T\to0$, it decreases rapidly as $e^{-E_{\textrm{a}}/T}$, where $E_{\textrm{a}}$ is the activation energy required to flip the occupation of the $f$-state at a single site. This thermally activated disordering of the chessboard phase also gives rise to the subgap states. As a result, the vertex contributions (dynamical charge screening) become negligible at low temperatures [for example, increasing the response in Fig.~\ref{fig: u05resp}(b)] and the total Raman response is determined by the bare loop contributions only. The expressions for the bare loop contributions [Eq.~(\ref{b1gP1}) for $B_{1g}$ symmetry and Eq.~(\ref{a1gP1}) for $A_{1g}$ symmetry] follow from Eqs.~(\ref{eq: Pi1X}) and (\ref{eq: chi0bar}) by choosing $\bm q=0$ ($X=1$) and $X'=0$ for the $B_{1g}$ symmetry and $X'=1$ for the $A_{1g}$ symmetry; the terms with $X^{\prime2}$ in Eq.~(\ref{eq: chi0bar}) determine the difference of the bare responses in the different symmetry channels.  From a mathematical standpoint, the presence of the additional terms due to a nonvanishing $X^\prime$, removes the singularity at the CDW gap edge which is in $\chi_0$.  From a physical standpoint, the different symmetries respond differently to the charge excitations.  The $A_{\rm 1g}$ channel responds with the full symmetry of the lattice, and the effect of the modulated CDW tends to average out this response so that the singular feature disappears, while this does not occur for the $B_{\rm 1g}$ channel, where a greatly enhanced response occurs near the charge gap. Hence, even though the vertex corrections vanish, which normally are required to guarantee that the system can screen long wavelength uniform charge fluctuations, the averaging effect of the CDW guarantees that the system can continue to screen these charge fluctuations even when the vertex corrections vanish.  For example, this is precisely how the uniform charge susceptibility will vanish in the CDW phase at $T=0$, which is required by the equations of motion for the total charge, and occurs due to the vertex corrections in the normal state above $T_c$; below $T_c$, since the vertex corrections are suppressed, this averaging feature takes over and allows the system to continue to effectively screen out those charge excitations.


In Fig.~{\ref{fig: u15resp}}, we plot the results for a near-critical Mott insulator with $U=1.5$.  The basic results remain quite similar to the metallic case.  We see the response function change dramatically as the system orders, with complex behavior at low temperature and low energy due to the subgap states, and then finally leading to the square root singularity in the $B_{\rm 1g}$ channel and smoother behavior in the $A_{\rm 1g}$ channel, with no singularity, and significantly reduced spectral weight.   The main change is the energy scale, since the gap is always identically equal to $U$ at $T=0$, and this is reflected in the ``pushing'' of the spectra to the right.  As we go from a near-gap insulator to a moderate-gap insulator with $U=2.5$ (Fig.~\ref{fig: u25resp}), we once again see similar kinds of behavior. In particular, we observe three peaks: the main CDW-gap peak at $\Omega=U$ is sharp for the $B_{1g}$ symmetry in panel (a) and smoothed for the $A_{1g}$ symmetry in panel $(b)$ and the two low-energy peaks have strong temperature dependence.

\begin{figure}[htb]
   \raisebox{-.45\height}{\includegraphics[scale=0.5]{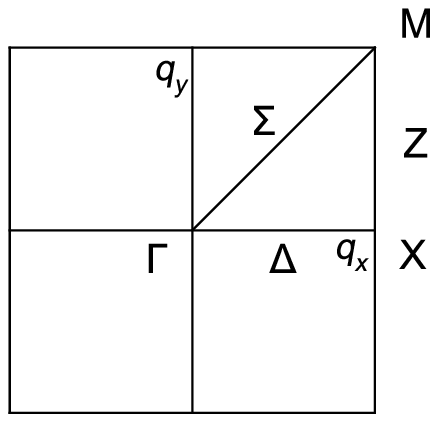}}
   \raisebox{-.45\height}{\includegraphics[scale=1.]{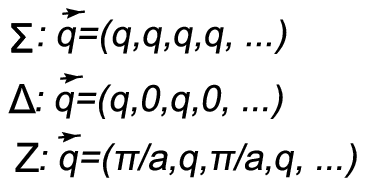}}
   \caption{Schematic of the first Brillouin zone with the high symmetry points labeled. Although we work in infinite
dimensions, we are trying to make contact with the two-dimensional BZ.}
   \label{fig: BZ}
\end{figure}

For nonresonant inelastic X-ray scattering, we investigate the behavior of the response functions  for the different transferred momentum values $\bm q$ in the first Brillouin zone (BZ). Because all the momentum dependence enters only through the parameters $X$ and $X'$, we must first understand their behavior in the BZ. We want our results to make contact with real physical systems, like a two-dimensional system, so we choose the following paths in the first BZ: the zone diagonal (zd) path lies in the so-called $\Sigma$-direction with $\bm q=(q,q,q,q,\ldots)$ and $-1\le X\le 1$; the zone edge (ze) path lies in the $Z$-direction with $\bm q=(\tfrac{\pi}{a},q,\tfrac{\pi}{a},q,\ldots)$ and $-1\le X\le 0$, and then continues along the zone edge path in the $\Delta$-direction with $\bm q=(q,0,q,0,\ldots)$ and $0\le X\le 1$.  These results are depicted in Fig.~\ref{fig: BZ}.
\begin{figure}[htb]
   \includegraphics[scale=0.6]{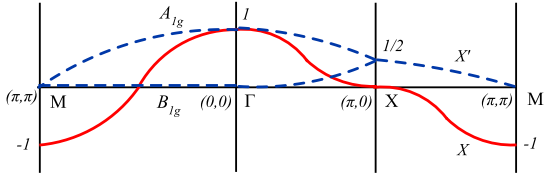}
   \caption{Plot of $X$ and $X'$ along the zone diagonal path and zone edge path in the first Brillouin zone.}
   \label{fig: XX'}
\end{figure}
The corresponding dependence of $X$ and $X'$ along these paths is plotted in Fig.~\ref{fig: XX'}. One can see, that along the $Z$-direction, the $X'$ value and, as a result, the response functions, are the same in both symmetry channels. For other directions, they are different. In addition, $X'=0$ along the zone diagonal $\Sigma$-direction for the $B_{1g}$ symmetry and the corresponding response function is determined only by the bare contributions with no vertex corrections (or dynamical screening) entering.

\begin{figure}[tb]
   \includegraphics[scale=0.8]{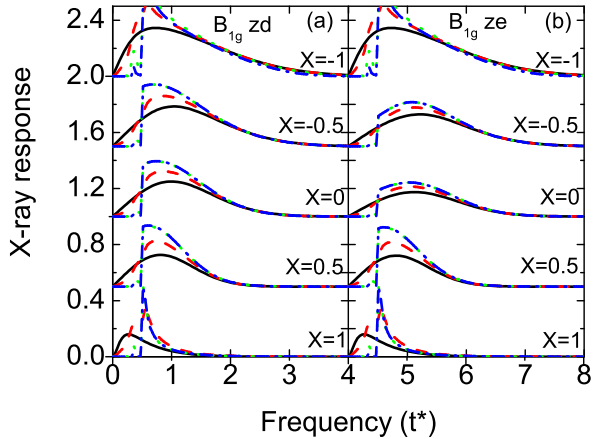}
   \caption{Nonresonant X-ray scattering  response function in the $B_{1g}$ channel for $U=0.5$ along the zone diagonal and zone edge of the first Brillouin zone. The set of curves correspond to temperatures $T=0.04$, $0.03$, $0.02$, and $0.015$.}
   \label{fig: u05b1gX}
\end{figure}

\begin{figure}[tb]
   \includegraphics[scale=0.8]{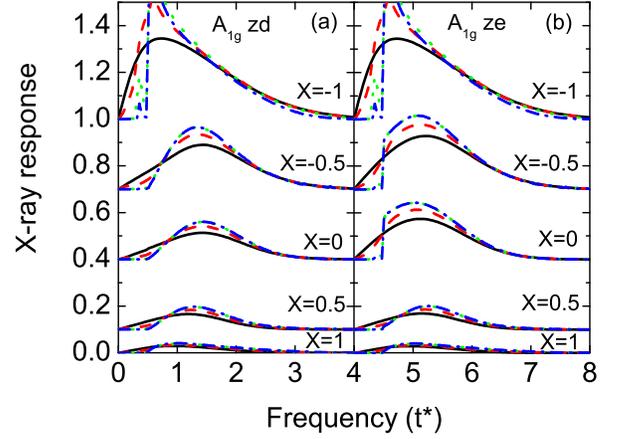}
   \caption{Nonresonant X-ray scattering  response function in the $A_{1g}$ channel for $U=0.5$ along the zone diagonal and zone edge of the first Brillouin zone. The set of curves correspond to temperatures $T=0.04$, $0.03$, $0.02$, and $0.015$.}
   \label{fig: u05a1gX}
\end{figure}

Having established the values of $X$ and $X'$ that we are using, we now show our numerical calculations of the nonresonant inelastic X-ray response functions for the case of a dirty metal with $U=0.5$ at different temperatures and transferred momentum. In Fig.~\ref{fig: u05b1gX}, we present results for the $B_{\rm 1g}$ symmetry and in Fig.~\ref{fig: u05a1gX} for the $A_{\rm 1g}$ symmetry. At the zone center ($X=1$), the response is the Raman scattering (see figures above) with sharp features in the $B_{\rm 1g}$ channel and with a strong suppression in the $A_{\rm 1g}$ channel. When we move away of the zone center, first of all, the sharp square root singularity at $\Omega=U$ in the $B_{\rm 1g}$ channel is rapidly replaced by a step-like response with a strong enhancement at the Brillouin zone corner $X=-1$, when the transferred momentum coincides with the CDW wave vector and we have effects of nesting present. For the $A_{\rm 1g}$ symmetry, we have a different scenario: there is a continuous enhancement without any sharp features, when we move along the zone diagonal and there is a continuous development of a step-like feature, when we move along the zone edge with a strong enhancement at the zone corner also due to nesting. In addition, the screening due to the vertex corrections vanishes there for all temperatures (since the $B_{\rm 1g}$ and $A_{\rm 1g}$ response functions are identical and have no vertex corrections there [$X'=0$]). In both cases, there is a large enhancement of the scattering  response function as we move from the zone center to the zone corner; this occurs because the system, as a whole, is much more effective at screening out uniform charge fluctuations than those modulated in space.
This shows, in particular, that scattering at the ordering wave vector is enhanced in the CDW system.



\begin{figure}[tb]
   \includegraphics[scale=0.8]{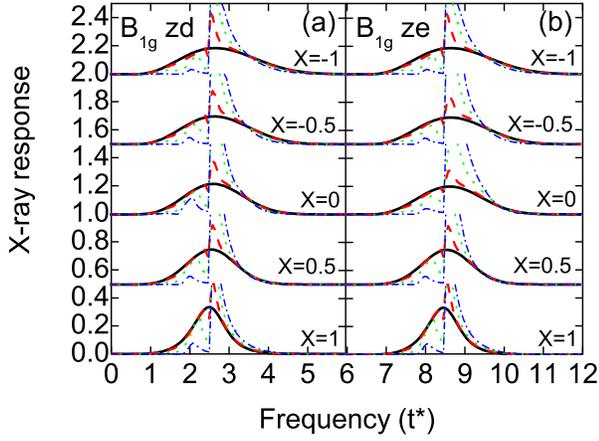}
   \caption{Nonresonant inelastic X-ray scattering  response function in the $B_{1g}$ channel for $U=2.5$ along the zone diagonal and the zone edge of the first BZ. The set of curves corresponds to the temperatures $T=0.08$, $0.07$, $0.06$,  and $0.04$.}
   \label{fig: u25b1gX}
\end{figure}

\begin{figure}[tb]
   \includegraphics[scale=0.8]{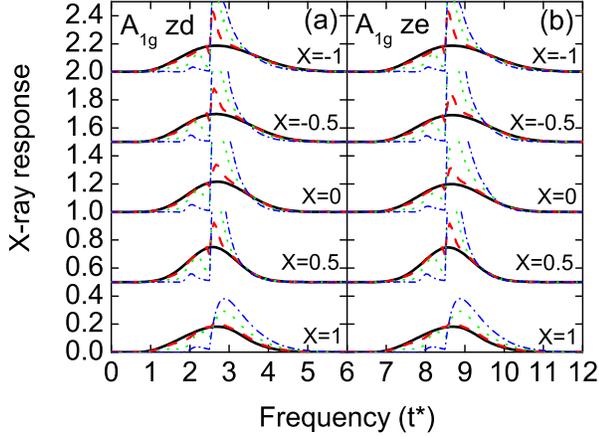}
   \caption{Nonresonant inelastic X-ray scattering  response function in the $A_{1g}$ channel for $U=2.5$ along the zone diagonal and the zone edge of the first BZ. The set of curves corresponds to the temperatures $T=0.08$, $0.07$, $0.06$, and $0.04$.}
   \label{fig: u25a1gX}
\end{figure}

Because the results for the near critical Mott insulator with $U=1.5$ are similar to the results for the other two $U$ values, we do not show them here.  But, we do plot the results for a small gap Mott insulator, with   $U=2.5$  in Figs.~\ref{fig:
u25b1gX} and \ref{fig: u25a1gX}). Here, we continue to see similar behavior to what is seen for $U=0.5$, namely, the character of the response changes rapidly as we move away from the zone-center, the differentiation of the results for different symmetry channels is reduced, and the results coincide at the zone corner.  We also see an enhancement of the signal and a generic broadening of the peaks as we move from the zone center to the zone corner.

Since we have derived first-moment sum rules for all of the response functions, we checked our numerical results by integrating the first moment of the response function and comparing that answer to the results of the moment sum rule expectation values, which are evaluated on the imaginary axis.  In all cases we examined, we achieved essentially perfect agreement, with errors less than 0.1\%, and arising primarily from the discretization we used in our frequency grid for the numerical integrations.

\begin{figure}[tb]
   \includegraphics[scale=0.7]{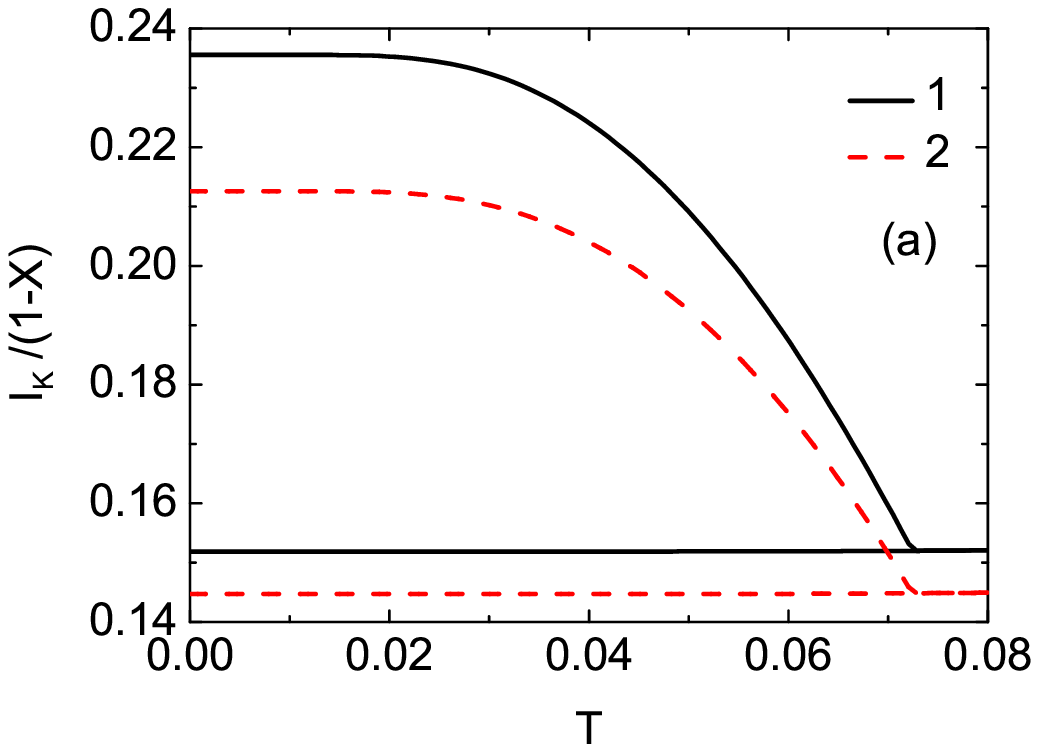}
   \includegraphics[scale=0.7]{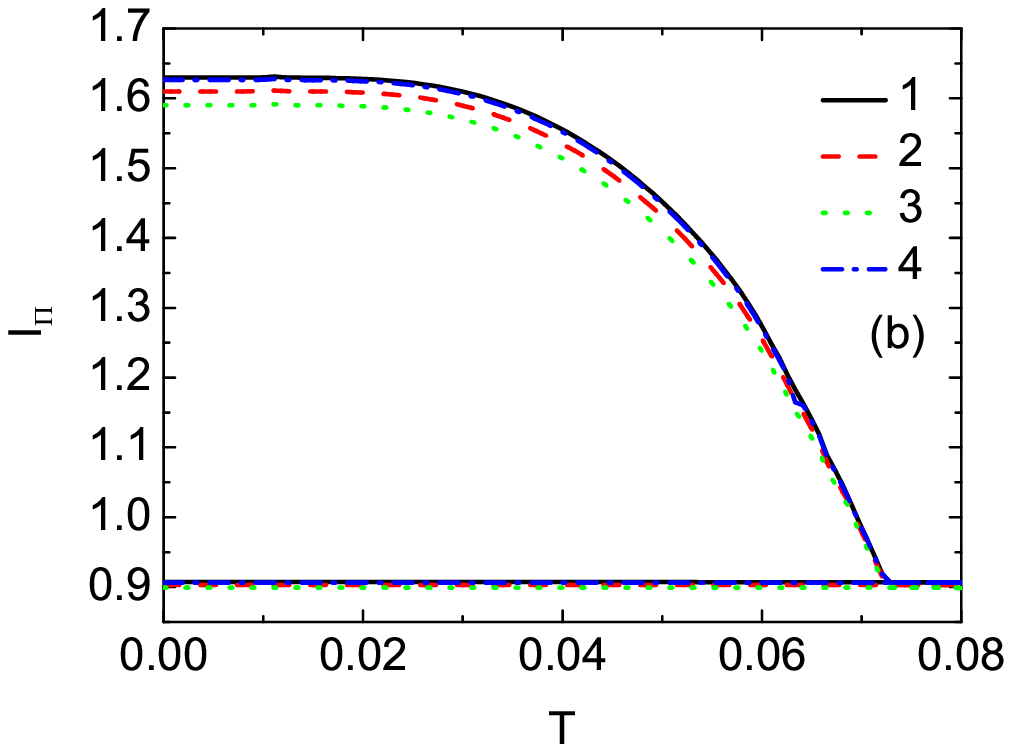}
   \caption{$B_{\rm 1g}$ sum rules as a function of temperature for $U=2.5$. (a) The kinetic-energy contribution  $I_K/(1-X)$: 1~--- zone-diagonal $\Sigma$-direction ($-1\le X \le1$, $X'=0$); 2~--- $X$-point at zone-edge ($X=0$). (b) The potential-energy contribution $I_{\Pi}$: 1~--- zone-diagonal $\Sigma$-direction ($-1\le X \le1$, $X'=0$); 2~--- zone-edge $X=-0.5$; 3~--- zone-edge $X=0$ ($X$-point); 4~--- zone-edge $X=0.5$. The thin lines correspond to the uniform solution artificially continued below $T_c$.}
   \label{fig: u25b1g_sumrule}
\end{figure}

\begin{figure}[tb]
   \includegraphics[scale=0.7]{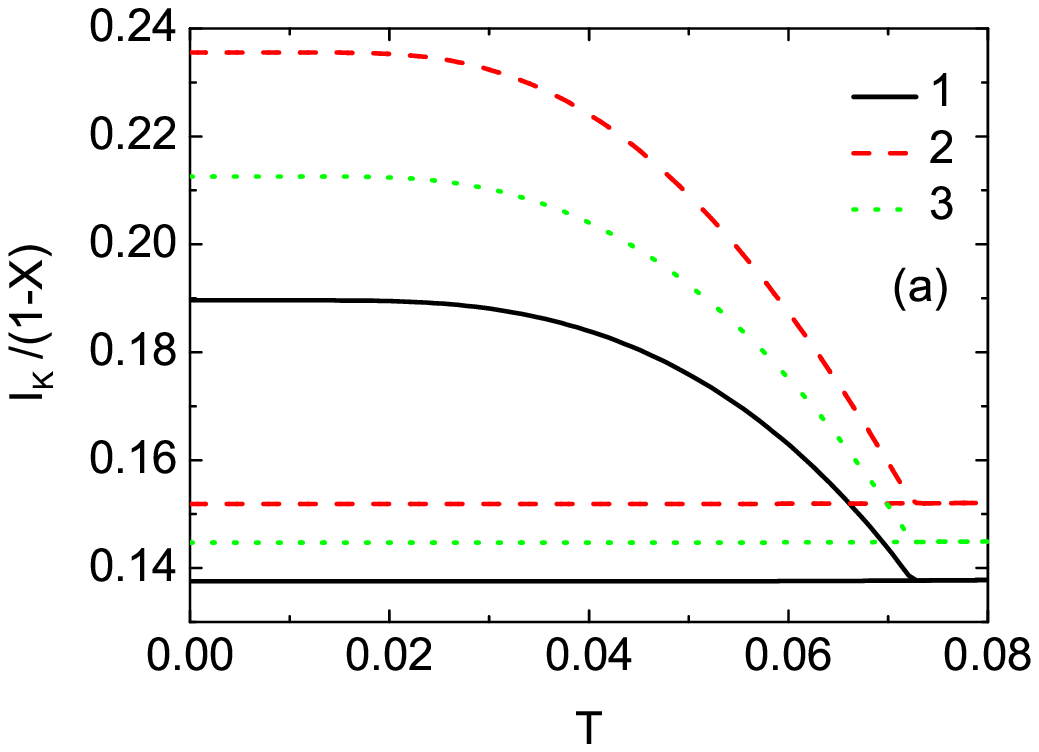}
   \includegraphics[scale=0.7]{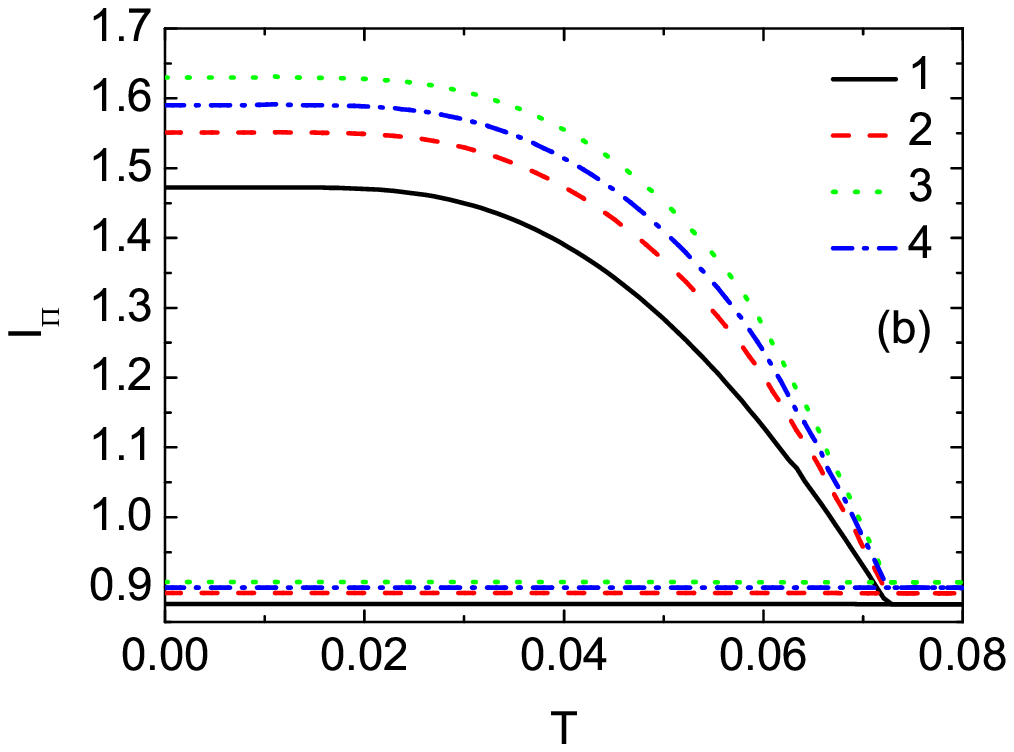}
   \caption{$A_{\rm 1g}$ sum rules as a function of temperature for $U=2.5$. (a) The kinetic-energy contribution $I_K/(1-X)$: 1~--- zone-diagonal $X=0$; 2~--- $M$-point at zone-corner ($X=-1$); 3~--- $X$-point at zone-edge ($X=0$). (b) The potential-energy contribution $I_{\Pi}$: 1~--- $\Gamma$-point at BZ center ($X=1$); 2~--- zone-diagonal $X=0$; 3~--- BZ corner $X=-1$ ($M$-point); 4~--- zone-edge $X=0$ ($X$-point). The thin lines correspond to the uniform solution artificially continued below $T_c$.}
   \label{fig: u25a1g_sumrule}
\end{figure}

\begin{figure}[tb]
   \includegraphics[scale=0.7]{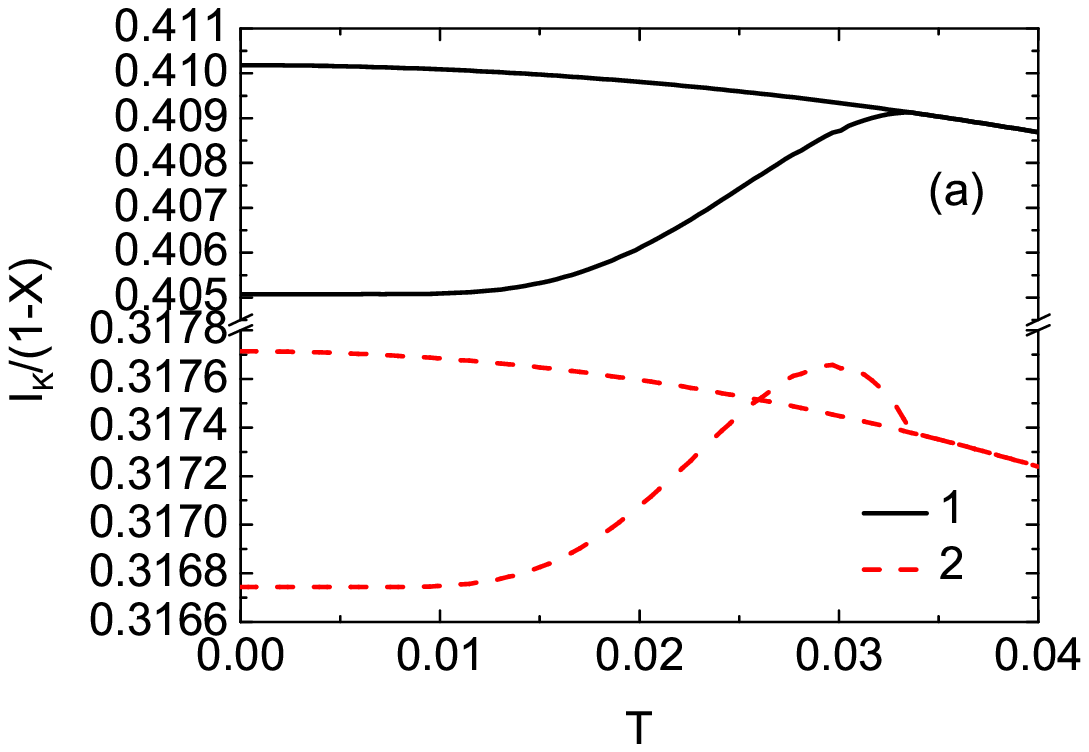}
   \includegraphics[scale=0.7]{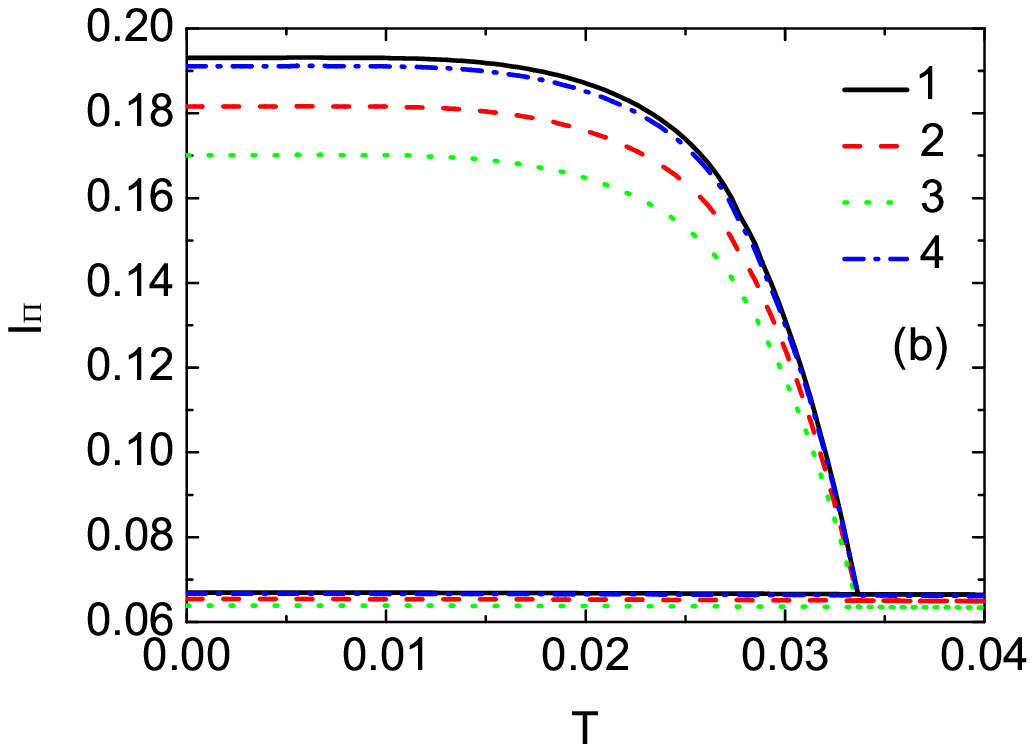}
   \caption{$B_{\rm 1g}$ sum rules as a function of temperature for $U=0.5$ (we plot the same cases as in Fig.~\ref{fig: u25b1g_sumrule}). }
   \label{fig: u05b1g_sumrule}
\end{figure}

\begin{figure}[tb]
   \includegraphics[scale=0.7]{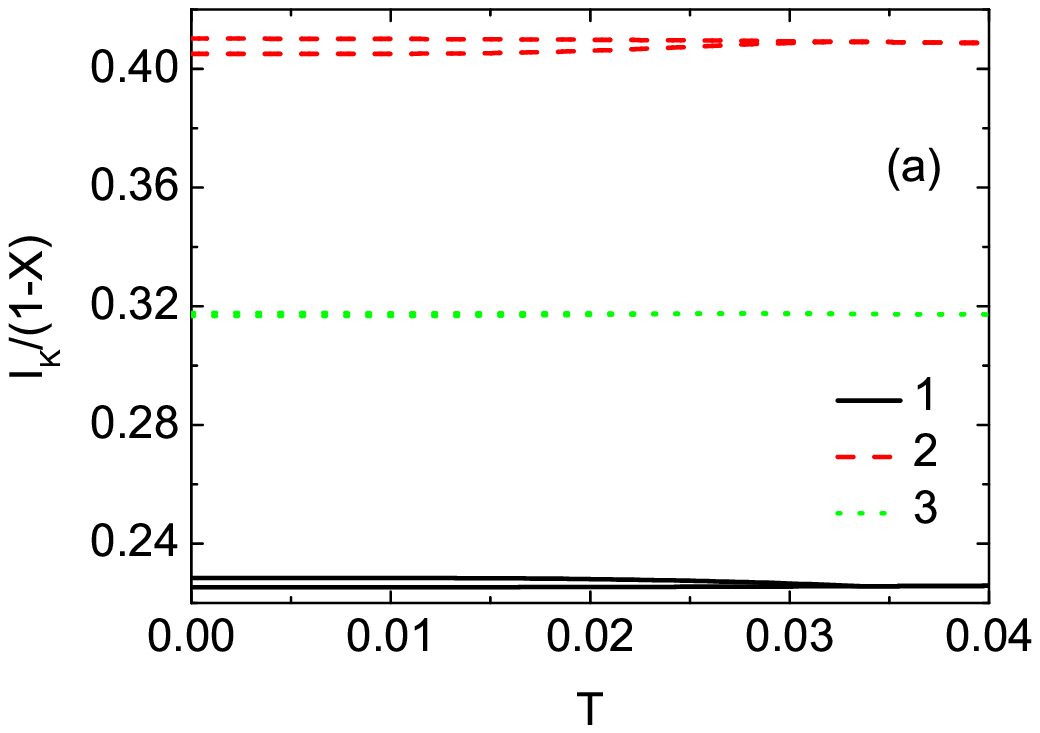}
   \includegraphics[scale=0.7]{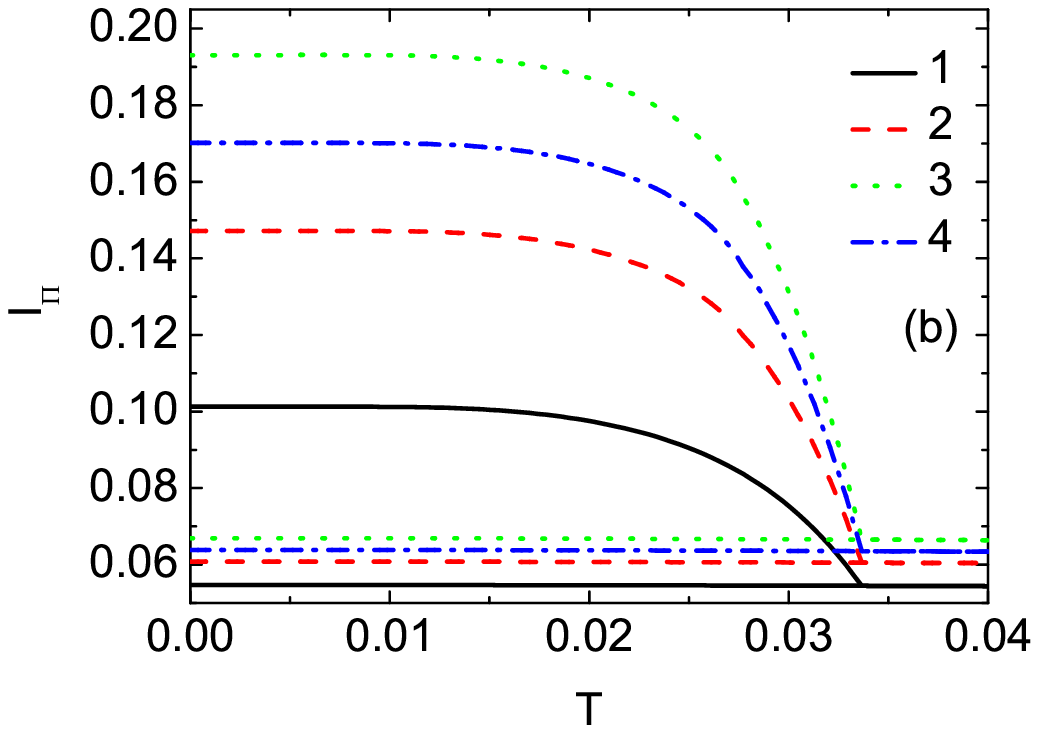}
   \caption{$A_{1g}$ sum rules as a function of temperature for $U=0.5$ (we plot the same cases as in Fig.~\ref{fig: u25a1g_sumrule}). }
   \label{fig: u05a1g_sumrule}
\end{figure}

But the sum rules can actually tell us more about the system.  One of the hallmarks of the $f$-sum rule for the optical conductivity is that the sum rule is fixed and does not change with temperature or interaction strength, so spectral weight is never lost or gained.  In a projected low-energy model, this result no longer holds, and the low-energy spectral weight can change with temperature or $U$, but, as is often the case, the changes are quite small at low temperature.  We can of course investigate this for our system in the CDW phase, by examining how the sum rule evolves for different parameters.
We begin with a plot of the sum rule for the case of strongly correlated insulator $U=2.5$ in the $B_{\rm 1g}$ channel in Fig.~\ref{fig: u25b1g_sumrule} and for the $A_{\rm 1g}$ channel in Fig.~\ref{fig: u25a1g_sumrule}. One can see, that for such values of Coulomb interaction the main contribution to the sum rule comes from the potential-energy part. The momentum dependence of the sum rule in the $B_{\rm 1g}$ channel is weak for the potential-energy contribution and strong for the kinetic-energy one [notice the $1-X$ factor in Eq.~(\ref{sum_rule_kinetic})]. For the $A_{\rm 1g}$ channel, both contributions have strong momentum dependence. For both symmetries, the largest values of the sum rule (total and for each contribution) are observed at the BZ corner $M$-point ($X=-1$) in both the uniform phase and the CDW phase, as could have been guessed due to the enhancement of the overall spectral functions we observed above (once again, in the CDW phase, we see an additional enhancement due to the ordering). The increase in the sum rule below $T_c$ is linear in $T_c-T$ and proportional to the square of the CDW order parameter $(\Delta n_f)^2$; this implies that if an experimental system has a nice separation between the low and high energy bands, then one could use this spectral weight to measure the order parameter as a function of temperature. For small values of $U$ (see Figs.~\ref{fig: u05b1g_sumrule} and \ref{fig: u05a1g_sumrule}), the kinetic-energy contribution gives the main contribution into the total sum rule. The kinetic-energy contribution continues to display strong momentum dependence and for some momentum its temperature dependence becomes quite nonlinear below $T_c$, as we already saw for the optical sum rule.\cite{matveev}

\section{Conclusions}

In this work, we developed the formalism (within DMFT) to calculate the nonresonant inelastic Raman and X-ray scattering in the case when the
system develops CDW order at low temperature.  The formalism is a straightforward generalization of the results
in the paramagnetic phase, but requires a careful accounting of the different sublattices and how they enter
into the diagrammatic expansions, and hence is technically quite challenging.  We also derived first-moment sum rules for these spectra and related the sum rules to different expectation values that can be immediately calculated.  We find that the sum rules relate to the potential energy in some cases, while in other cases, both the kinetic energy and the potential energy terms enter into the expectation values (and also the CDW order parameter).

We applied our formalism to the case of the spinless Falicov-Kimball model because the charge vertex is known exactly for that system, and hence we can find an exact numerical solution to the light scattering response functions.
The main numerical result that we find is that there is very strong temperature dependence that sets in once we pass through $T_c$.  This occurs because the system rapidly depletes subgap states as it forms the CDW gap, and then develops a square-root singularity due to the pile-up of states at $T=0$.  These features can be immediately seen in the light scattering response functions, but are symmetry selective. When vertex corrections act to screen the light scattering at high temperatures (near and above $T_c$), the square root singularity is suppressed in the $A_{\rm 1g}$ channel, as is the overall magnitude of the light scattering signal. At low temperatures, the most important effects are due to nesting of the transferred momentum in BZ and due to an effective screening of uniform charge fluctuations which arises due to an averaging effect over the modulated charge distribution of the system. The qualitative shape of the response function for inelastic X-ray scattering, where the spectra has almost a discontinuous jump near the gap, is an unexpected result, that occurs when one combines the square-root singularity associated with the pile-up of the density of states near the gap edge with the nontrivial nesting effects and the dynamical charge screening effects of the many-body system. While we see an enhancement of the response, a broadening of the spectra, and an increase in the magnitude of the sum rule as we move from the zone center to the zone corner, we do not see any dramatic changes in the shape of the spectra associated with the fact that we can transfer momenta that is equal to the ordering wave vector of the CDW but we do see a significant overall enhancement of the signal.  This turns out to be similar to what was seen in the dynamical charge susceptibility of the model as one approaches $T_c$ from above\cite{charge_vertex}, and may be related to the fact that the Falicov-Kimball model has a reducible charge vertex that assumes very different behavior for dynamical charge fluctuations as it does for static charge fluctuations, which give rise to the underlying CDW order. If true, then we would anticipate even larger effects in  models where the charge vertex is not decoupled in this fashion, such as the Hubbard or Holstein model, but resolving this question is a problem for the future.

Our numerical work focused on the case of half filling.  One might ask what would happen away from half-filling.  While it is true that the CDW phase can be seen as the first ordered phase as we go from the normal state to an ordered state at $T_c$,\cite{freericks_14797} we do not know whether the AB ordered phase survives all the way to $T=0$ or whether there are subsequent phase transitions, perhaps to incommensurate phases as $T$ is further reduced.  For this reason, we have not chosen to solve such problems in this work.  In the high-temperature phases, where the system is ordered in the AB CDW, the chemical potential would need to be located outside of the gap, and so we would expect to see more response at low energies, but as the $T$ was further lowered, we expect incommensurate order to enter, and for the system to have a well developed gap, so that the results would most likely look similar to those shown here. On the other hand, there is another possible scenario at low temperatures for some densities of the mobile and localized electrons when, instead of the incommensurate order, the phase separation into chess-board and uniform phases can take place.\cite{freericks_lemanski,stasyuk} In this case the total response will be a sum of the responses for the chess-board and uniform phases weighted by the volume fractions of these phases.

We believe our results will be most relevant to electronic Raman or X-ray scattering on CDW ordered systems in three dimensions.  So far, most of the Raman scattering work has focused on understanding how phonons behave as one passes through the transition, including the behavior of the phonon softening for the CDW mode\cite{cooper}.  We hope that our results will inspire experimental groups to also consider examining electronic Raman scattering in CDW systems to see whether they display the kinds of features that we showed here.

In the future, we will generalize the resonant light scattering formalism to the CDW phase and examine what modifications enter into the response functions in that case.

\begin{acknowledgements}
This publication is based on work supported by Award No. UKP2-2697-LV-06 of the U.S. Civilian Research and Development Foundation (CRDF).
J.~K.~F.~acknowledges support from the Department of Energy, Office of Basic Science, under Grants No.~DE-FG02-08ER46540 (for the collaroration) and No.~DE-FG02-08ER46542. 
We would also like to acknowledge the comments of an anonymous referee who led us to a better understanding of how screening occurs in the low-temperature regime.
\end{acknowledgements}


\appendix*

\section{Sum rule derivation}

In this appendix, we present details for the derivation of the first-moment sum rules of inelastic light and X-ray scattering in the ordered CDW phase. To begin, we must evaluate the first
commutator in Eq.~(\ref{sum_rule_def}) which yields
\begin{align}
  \left[H,\tilde\gamma(\bm q)\right]&=
  U\sum_{ija} t_{ij}^{a\bar a} e^{i\bm Q(\bm R_i^a-\bm R_j^{\bar a})}
  e^{-i\frac{\bm q}2 (\bm R_i^a+\bm R_j^{\bar a})}
  \\
  &\times\left(\hat n_{if}^a -\hat n_{jf}^{\bar a} \right)
  \hat d_{ia}^{\dag}\hat d_{j\bar a}
  \nonumber\\
  &-\sum_{ijla} t_{il}^{a\bar a} t_{lj}^{\bar aa} \left[
  e^{i\bm Q(\bm R_l^{\bar a}-\bm R_j^a)} e^{-i\frac{\bm q}2 (\bm R_l^{\bar a}+\bm R_j^a)}
  \right.
  \nonumber\\
  &-\left.
  e^{i\bm Q(\bm R_i^a-\bm R_l^{\bar a})} e^{-i\frac{\bm q}2 (\bm R_i^a+\bm R_l^{\bar a})} \right]
  \hat d_{ia}^{\dag}\hat d_{ja}.
  \nonumber
\end{align}
Here we introduce the notations $\bar A=B$ and $\bar B=A$ and use the fact that the hopping integral connects only sites which belong to different sublattices.

\begin{widetext}
The second commutator now gives
\begin{align}
  \left[\gamma^{\dag}(\bm q)\left[H,\tilde\gamma(\bm q)\right]\right]&=
  U\sum_{ijla} t_{il}^{a\bar a} t_{lj}^{\bar aa} e^{i\bm Q(\bm R_i^a-\bm R_j^a)} \left[
  e^{i\frac{\bm q}2 (\bm R_i^a-\bm R_j^a)} \left(\hat n_{lf}^{\bar a} -\hat n_{jf}^a \right) -
  e^{-i\frac{\bm q}2 (\bm R_i^a-\bm R_j^a)} \left(\hat n_{if}^a -\hat n_{lf}^{\bar a} \right) \right]
  \hat d_{ia}^{\dag}\hat d_{ja}
  \\
  &-\sum_{ijlna} t_{il}^{a\bar a} t_{ln}^{\bar aa} t_{nj}^{a\bar a} \left[
  e^{i\bm Q(\bm R_i^a - \bm R_l^{\bar a} + \bm R_n^a - \bm R_j^{\bar a})}
  e^{i\frac{\bm q}2 (\bm R_i^a + \bm R_l^{\bar a} - \bm R_n^a - \bm R_j^{\bar a})} -
  e^{i\bm Q(\bm R_i^a-\bm R_n^a)} e^{i\frac{\bm q}2 (\bm R_i^a-\bm R_n^a)} \right.
  \nonumber\\
  &+\left.
  e^{i\bm Q(\bm R_i^a - \bm R_l^{\bar a} + \bm R_n^a - \bm R_j^{\bar a})}
  e^{-i\frac{\bm q}2 (\bm R_i^a + \bm R_l^{\bar a} - \bm R_n^a - \bm R_j^{\bar a})} -
  e^{i\bm Q(\bm R_l^{\bar a}-\bm R_j^{\bar a})} e^{-i\frac{\bm q}2 (\bm R_l^{\bar a}-\bm R_j^{\bar a})}
  \right]\hat d_{ia}^{\dag}\hat d_{j\bar a}.
  \nonumber
\end{align}
Next, we use the fact that the hopping is allowed only between NN sites, and we replace $j=i+\delta$ in $t_{ij}^{ab}$, where $\delta$ runs over all of the NNs of site $i$, to obtain
\begin{align}\label{eq: sumrule3}
  \left\langle\left[\gamma^{\dag}(\bm q)\left[H,\tilde\gamma(\bm q)\right]\right]\right\rangle&=
  Ut^2\sum_{i\delta\delta'a} e^{-i\bm Q(\bm\delta+\bm\delta')} \left\langle\left[
  e^{-i\frac{\bm q}2 (\bm\delta+\bm\delta')}
  \left(\hat n_{i+\delta,f}^{\bar a} -\hat n_{i+\delta+\delta',f}^a \right) -
  e^{i\frac{\bm q}2 (\bm\delta+\bm\delta')} \left(\hat n_{i,f}^a -\hat n_{i+\delta,f}^{\bar a} \right)
  \right] \hat d_{i,a}^{\dag} \hat d_{i+\delta+\delta',a} \right\rangle
  \\
  &-t^3\sum_{i\delta\delta'\delta''a} e^{-i\bm Q(\bm\delta'+\bm\delta'')} \left[
  e^{-i\frac{\bm q}2 (\bm\delta'+\bm\delta'')}\left(e^{-i\bm q\bm\delta}-1\right) +
  e^{i\frac{\bm q}2 (\bm\delta'+\bm\delta'')}\left(e^{i\bm q\bm\delta}-1\right)
  \right] \left\langle\hat d_{i,a}^{\dag}\hat d_{i+\delta+\delta'+\delta'',\bar a} \right\rangle.
  \nonumber
\end{align}
The first term contains expectation values of three operator products which can be calculated by introducing an auxiliary field $\mu_f^c\to \mu_f^c + \delta\mu_{l,f}^c$ at site $l$ into the Hamiltonian and taking a functional derivative
\begin{align}\label{eq: 3op}
  \left\langle\hat d_{jb}^{\dag}\hat d_{ia}\hat n_{l,f}^c \right\rangle & = T\sum_m \left[
  T\frac{\delta G_{ij}^{ab}(i\omega_m)}{\delta\mu_{l,f}^c} + n_{l,f}^c G_{ij}^{ab}(i\omega_m)\right]
  = T\sum_m \left[ G_{il}^{ac}(i\omega_m)
  T\frac{\delta \Sigma_{l}^{c}(i\omega_m)}{\delta\mu_{l,f}^c} G_{lj}^{cb}(i\omega_m) +
  n_{l,f}^c G_{ij}^{ab}(i\omega_m)\right].
\end{align}
One can immediately calculate the derivative
\begin{equation}\label{eq: dSigma}
  T\frac{\delta \Sigma_{l}^{c}(i\omega_m)}{\delta\mu_{l,f}^c}=
  \frac{1}{(G_m^c)^2}\;\frac{U n_f^c(1- n_f^c)}
  {(i\omega_m+\mu_d^c-\lambda_m^c)(i\omega_m+\mu_d^c-\lambda_m^c-U)}
  =\frac{\Sigma_m^c-U n_f^c}{UG_m^c}
\end{equation}
from the solution of the single-impurity problem. After substituting this result into Eq.~(\ref{eq: sumrule3}), we find that
\begin{align}
  I &= - \frac{\pi T}{2} \sum_m \sum_a
  \frac{1}{N}\sum_{\bm k} G_{\bm k}^{\bar aa}(i\omega_m)
  \left[\epsilon_{\bm k-\frac{\bm q}{2}-\bm Q}^2(\epsilon_{\bm k-\bm q} -\epsilon_{\bm k}) +
  \epsilon_{\bm k+\frac{\bm q}{2}-\bm Q}^2(\epsilon_{\bm k+\bm q} -\epsilon_{\bm k})\right]
  \\
  &+ \frac{\pi T}{2} \sum_m \sum_a  \frac{\Sigma_m^a-U n_f^a}{G_m^a} \left\{
  \frac{1}{N}\sum_{\bm k} G_{\bm k}^{a\bar a}(i\omega_m) \epsilon_{\bm k-\frac{\bm q}{2}-\bm Q}
  \frac{1}{N}\sum_{\bm k'} G_{\bm k'}^{\bar aa}(i\omega_m) \epsilon_{\bm k'-\frac{\bm q}{2}-\bm Q}
  \right.
  \nonumber\\
  &+\left.
  \frac{1}{N}\sum_{\bm k} G_{\bm k}^{a\bar a}(i\omega_m) \epsilon_{\bm k+\frac{\bm q}{2}-\bm Q}
  \frac{1}{N}\sum_{\bm k'} G_{\bm k'}^{\bar aa}(i\omega_m) \epsilon_{\bm k'+\frac{\bm q}{2}-\bm Q}
  \right\}
  \nonumber\\
  &-\frac{\pi T}{2} \sum_m \sum_a \left[ \Sigma_m^a-U n_f^a + U(n_f^a- n_f^{\bar a})\right]
  \frac{1}{N}\sum_{\bm k} G_{\bm k}^{aa}(i\omega_m)
  \left(\epsilon_{\bm k-\frac{\bm q}{2}-\bm Q}^2 + \epsilon_{\bm k+\frac{\bm q}{2}-\bm Q}^2\right).
  \nonumber
\end{align}
The summations over momentum can be explicitly performed as follows:
\begin{align}
  &\frac{1}{N}\sum_{\bm k} G_{\bm k}^{a\bar a}(i\omega_m) \epsilon_{\bm k-\frac{\bm q}{2}-\bm Q}
  = X'\left[ \bar Z_m F_\infty(\bar{Z}_m) -1 \right];
  \\
  &\frac{1}{N}\sum_{\bm k} G_{\bm k}^{aa}(i\omega_m) \epsilon_{\bm k-\frac{\bm q}{2}-\bm Q}^2
  = \frac{Z_m^{\bar a}}{\bar Z_m}\left\{ \frac{t^{*2}}{2} (1-X^{\prime2}) F_\infty(\bar{Z}_m)
  + X^{\prime2} \bar Z_m \left[ \bar Z_m F_\infty(\bar{Z}_m) -1 \right] \right\};
  \nonumber\\
  &\frac{1}{N}\sum_{\bm k} G_{\bm k}^{\bar aa}(i\omega_m)
  \epsilon_{\bm k-\frac{\bm q}{2}-\bm Q}^2(\epsilon_{\bm k-\bm q} -\epsilon_{\bm k})=
  \frac{t^{*2}}{2}(1-X)\left[ \bar Z_m F_\infty(\bar{Z}_m) -1 \right]
  \nonumber\\
  &- X^{\prime2}(1-X)\left[ \frac32 t^{*2} \left[ \bar Z_m F_\infty(\bar{Z}_m) -1 \right] -
  \bar Z_m^2 \left[ \bar Z_m F_\infty(\bar{Z}_m) -1 \right] + \frac{t^{*2}}{2} \right];
  \nonumber
\end{align}
where
\begin{equation}
  Z_m^a=i\omega_m+\mu_d^a-\Sigma_m^a.
\end{equation}

Finally, the sum rule (first moment of the response function) contains two contributions
\begin{equation}
  I = I_K + I_{\Pi}.
\end{equation}
The first contribution comes from the kinetic energy term
\begin{align}
  I_K &= 2(1-X)\int_{-\infty}^{+\infty} d\omega f(\omega)\Img\left\{
  \frac{t^{*2}}{2}\left[ \bar Z(\omega) F_\infty[\bar{Z}(\omega)] -1 \right]
  \right.
  \\
  &\left.
  - X^{\prime2}\left[ \frac32 t^{*2} \left[ \bar Z(\omega) F_\infty[\bar{Z}(\omega)] -1 \right] -
  \bar Z^2(\omega) \left[ \bar Z(\omega) F_\infty[\bar{Z}(\omega)] -1 \right] + \frac{t^{*2}}{2} \right]
  \right\}
  \nonumber
\end{align}
and the second one comes from the potential-energy term
\begin{align}
  I_{\Pi} &= \!\int_{-\infty}^{+\infty}\! d\omega f(\omega)\Img\sum_a\Biggl\{
  \left[ \Sigma^a(\omega) -U n_f^a \right] \Biggl(
  \frac{t^{*2}}{2}(1-X^{\prime2})G^a(\omega)
  +X^{\prime2}Z^{\bar a}(\omega)\left[ \bar Z(\omega) F_\infty[\bar{Z}(\omega)] -1 \right]
\\
  &-X^{\prime2}\frac{\left[ \bar Z(\omega) F_\infty[\bar{Z}(\omega)] -1 \right]^2}{G^a(\omega)}
  \Biggr)
  +U(n_f^a- n_f^{\bar a}) \left( \frac{t^{*2}}{2}(1-X^{\prime2})G^{aa}(\omega)
  + X^{\prime2}Z^{\bar a}(\omega)\left[ \bar Z(\omega) F_\infty[\bar{Z}(\omega)] -1 \right] \right)
\Biggr\}.
\nonumber
\end{align}
\end{widetext}
Using the identities $\bar Z(\omega)F_\infty[\bar{Z}(\omega)]-1 = \lambda^A(\omega)G^{AA}(\omega) = \lambda^B(\omega)G^{BB}(\omega)$ and $[G^{aa}(\omega)]^{-1}=Z^a(\omega)-\lambda^a(\omega)$, we can rewrite the potential-energy term contribution in the final form of Eq.~(\ref{sum_rule_potential}), where we use the fact that in equilibrium $\mu_d^A=\mu_d^B$ and $Z^A(\omega)-Z^B(\omega)=\Sigma^B(\omega)-\Sigma^A(\omega)$.

\end{document}